\pdfoutput=1 

\documentclass[12pt]{article}

\usepackage[height=25cm,a4paper,hmargin={2.8cm,2.2cm}]{geometry}

\usepackage{amsmath,amssymb}
\usepackage{graphicx}
\usepackage[colorlinks=true,linkcolor=blue,urlcolor=blue,citecolor=blue]{hyperref}
\usepackage{cellspace}

\usepackage{slashed}

\usepackage[square,comma,sort&compress,numbers]{natbib}

\usepackage{caption}
\usepackage{authblk}

\renewcommand{\theequation}{\thesection.\arabic{equation}}

\DeclareMathOperator{\tr}{tr} 
\DeclareMathOperator{\re}{Re} 
\DeclareMathOperator{\dilog}{Li_2} 

\begin{document}

\title{A new event generator for the elastic scattering of~charged leptons on protons}
\author[1]{A.\,V.~Gramolin\thanks{E-mail:~\href{mailto:gramolin@inp.nsk.su}{gramolin@inp.nsk.su}}}
\author[1,2]{V.\,S.~Fadin}
\author[1,2]{A.\,L.~Feldman}
\author[1,2]{R.\,E.~Gerasimov}
\author[1]{D.\,M.~Nikolenko}
\author[1]{I.\,A.~Rachek}
\author[1,2]{D.\,K.~Toporkov}
\affil[1]{Budker Institute of Nuclear Physics, 630090 Novosibirsk, Russia}
\affil[2]{Novosibirsk State University, 630090 Novosibirsk, Russia}
\date{}

\maketitle

\begin{abstract}
This paper describes a new multipurpose event generator, \texttt{ESEPP}, which has been developed for the Monte Carlo simulation of unpolarized elastic scattering of charged leptons on protons. The generator takes into account the lowest-order QED radiative corrections to the Rosenbluth cross section including first-order bremsstrahlung without using the soft-photon or ultrarelativistic approximations. \texttt{ESEPP} can be useful for several significant ongoing and planned experiments.

\bigskip
\noindent
Keywords: event generator, elastic lepton--proton scattering, radiative corrections, proton electromagnetic form factors, two-photon exchange
\end{abstract}

\section{Introduction and motivation}

\begin{figure}[t]
\centering
\includegraphics[width=0.8\textwidth]{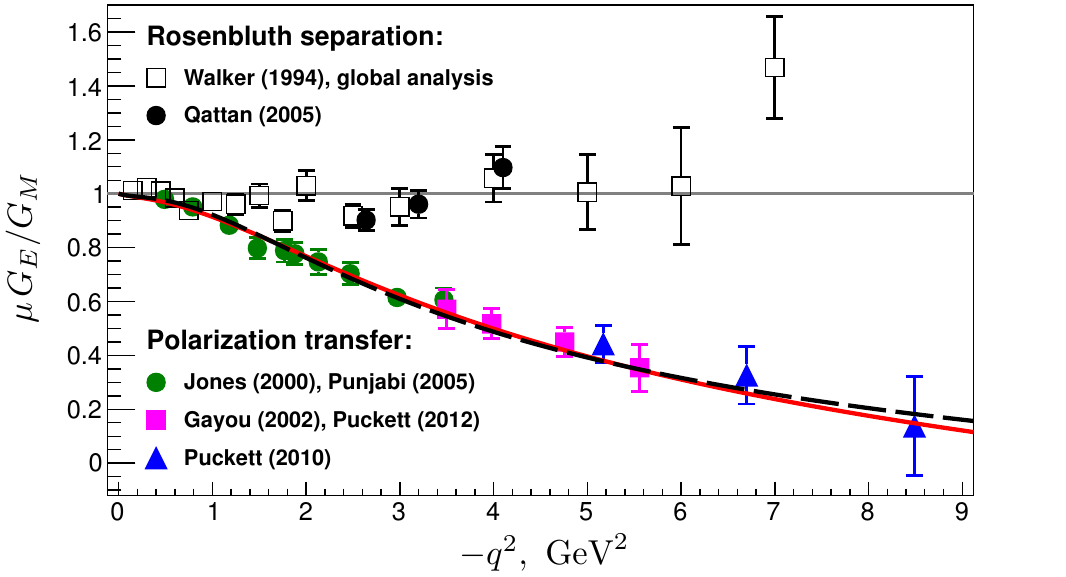}
\caption{Comparison of the experimental data for the ratio~$\mu \, G_E / G_M$ obtained by the Rosenbluth separation method~\cite{PRD.49.5671, PRL.94.142301} and polarization transfer method~\cite{PRL.84.1398, PRC.71.055202, PRL.88.092301, PRC.85.045203, PRL.104.242301}. Here $\mu$~is the magnetic moment of the proton. The curves correspond to the parametrizations of the proton form factors by Kelly~\cite{PRC.70.068202} (black dashed line) and Puckett~\cite{arXiv:1008.0855} (red solid line).}
\label{Fig1}
\end{figure}

During the last decade, a renewed interest has been shown in elastic electron--proton scattering and the proton's electric and magnetic form factors, $G_E (q^2)$ and $G_M (q^2)$, in the spacelike region, where $q^2 < 0$. The main reason for this interest is that the recent data~\cite{PRL.84.1398, PRC.71.055202, PRL.88.092301, PRC.85.045203, PRL.104.242301} for the ratio~$G_E / G_M$ obtained with polarization methods disagree with the older results for the proton form factors obtained with the Rosenbluth separation technique (see the review~\cite{PPNP.59.694}). The discrepancy between the two sets of data for the form factor ratio is clearly seen in figure~\ref{Fig1}. In addition, there is a recently-observed contradiction between the value of the proton charge radius extracted from the Lamb shift in muonic hydrogen~\cite{Nature.466.213, Science.339.417} and the value measured using elastic electron--proton scattering and atomic hydrogen spectroscopy~\cite{RMP.80.633, PRL.105.242001, PLB.705.59, ARNPS.63.175}.

This interest is the reason why several new experiments with unpolarized elastic lepton--proton scattering are currently in preparation or have already been carried out in recent years. First, we would like to mention the experiments aimed at studying the two-photon exchange (TPE) effects in elastic scattering of electrons and positrons on protons. Unaccounted TPE corrections are considered to be the most likely cause of the above-mentioned discrepancy between two methods of measuring the proton form factors (see the review papers \cite{ARNPS.57.171, PPNP.66.782}). The most direct way of studying the TPE effects is by a precise comparison of the cross sections for elastic $e^- p$ and $e^+ p$ scattering. Such measurements were recently carried out by the three collaborations: \mbox{VEPP--3} at~BINP (Novosibirsk, Russia) \cite{nucl-ex.0408020, NPBPS.225-227.216}, OLYMPUS at~DESY (Hamburg, Germany) \cite{AIPConfProc.1160.19, NIMA.741.1}, and CLAS at~JLab (Newport News, VA, USA)~\cite{AIPConfProc.1160.24, PRC.88.025210}. These three experiments have already finished data collection and are now in the analysis stage. An important part of this stage is to accurately take into account the standard radiative corrections, especially bremsstrahlung, because it leads to a difference between the cross sections of elastic $e^- p$ and $e^+ p$ scattering.

The size of the radiative corrections depends on the type of the detector used (magnetic or non-magnetic), the detector acceptance, its spatial/angular and energy/momentum resolutions, and the kinematic cuts applied to select elastic scattering events. For this reason, a careful account of the radiative corrections in each of these experiments requires a realistic Monte Carlo simulation of the detector response, for example, using the Geant4 toolkit. This in turn requires an event generator which outputs kinematic parameters of all final-state particles. We have developed such a generator and named it \texttt{ESEPP} (Elastic Scattering of Electrons and Positrons on Protons). \texttt{ESEPP} takes into account the first-order quantum electrodynamics (QED) radiative corrections to the Rosenbluth cross section of unpolarized elastic scattering of charged leptons ($e^{\pm}$ or $\mu^{\pm}$) on protons. Weak interaction effects are neglected in this study. Particular attention is paid to first-order bremsstrahlung, which we consider without using the soft-photon or ultrarelativistic approximations. Note that throughout the paper we discuss only so-called internal bremsstrahlung, where the photon is emitted during the act of hard lepton--proton scattering. One should also remember about the processes accompanying the passage of the incident and outgoing particles through the target and detector materials, the vacuum chamber, the air gaps, etc.

\texttt{ESEPP} may also be useful for two upcoming experiments aimed at studying the proton charge radius with elastic lepton--proton scattering: the \mbox{E12-11-106} experiment~\cite{PRad, EPJWebConf.73.07006}, being prepared by the PRad collaboration at~JLab, and the muon--proton scattering experiment~\cite{arXiv:1303.2160}, proposed by the MUSE collaboration at~PSI (Villigen, Switzerland). There are two advantages of the generator making it suitable for these experiments. First, \texttt{ESEPP} does not use the common approximation $-q^2 \gg m^2$ (where $m$~is the lepton mass) in the calculation of first-order bremsstrahlung. Second, it implements an accurate (without neglecting the lepton mass) calculation of the event kinematics, which is a necessary requirement for the case of non-ultrarelativistic muons.

\begin{figure}[t]
\centering
\begin{tabular}{ccccc}
\includegraphics[width=0.15\textwidth]{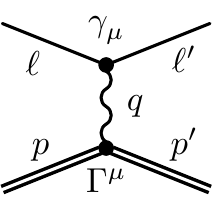} &
\includegraphics[width=0.15\textwidth]{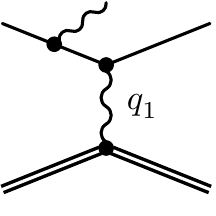} &
\includegraphics[width=0.15\textwidth]{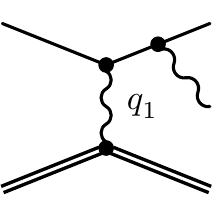} &
\includegraphics[width=0.15\textwidth]{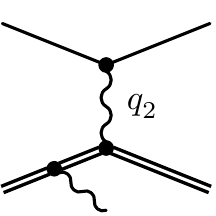} &
\includegraphics[width=0.15\textwidth]{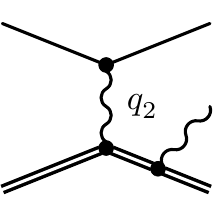} \\
(a) ${\mathcal M}_{\text{Born}}$ & (b) ${\mathcal M}_{\text{brems}}^{\text{li}}$ & (c) ${\mathcal M}_{\text{brems}}^{\text{lf}}$ & (d) ${\mathcal M}_{\text{brems}}^{\text{pi}}$ & (e) ${\mathcal M}_{\text{brems}}^{\text{pf}}$ \\
& & & & \\
\includegraphics[width=0.15\textwidth]{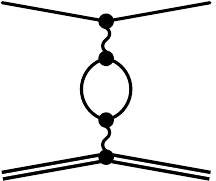} &
\includegraphics[width=0.15\textwidth]{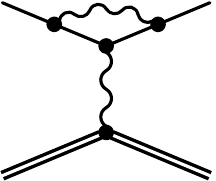} &
\includegraphics[width=0.15\textwidth]{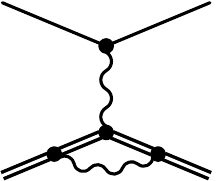} &
\includegraphics[width=0.15\textwidth]{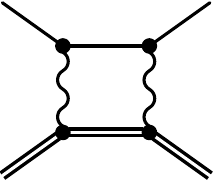} &
\includegraphics[width=0.15\textwidth]{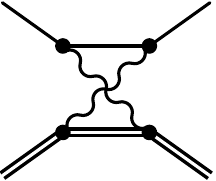} \\
(f) ${\mathcal M}_{\text{vac}}$ & (g) ${\mathcal M}_{\text{vert}}^{\ell}$ & (h) ${\mathcal M}_{\text{vert}}^{p}$ & (i) ${\mathcal M}_{\text{box}}$ & (j) ${\mathcal M}_{\text{xbox}}$
\end{tabular}
\caption{Feynman diagrams representing the elastic scattering of charged leptons on protons in the leading and next-to-leading orders. (a) corresponds to the one-photon exchange or first Born approximation. (b)--(e) show the first-order bremsstrahlung process $\ell^{\pm} p \rightarrow \ell^{\pm} p \, \gamma$ in the cases when the photon is emitted by the initial-state lepton~(b), final-state lepton~(c), initial-state proton~(d), or final-state proton~(e). (f)--(j) represent the processes contributing to the virtual-photon corrections: the vacuum polarization correction~(f), the lepton~(g) and proton~(h) vertex corrections, and the TPE corrections (i), (j).}
\label{Fig2}
\end{figure}

Outside of the study of proton form factors, \texttt{ESEPP} can be used for the Monte Carlo simulation of the two proposed experiments~\cite{arXiv:1207.5089, PRD.86.115012, arXiv:1307.4432} searching for a hypothetical light gauge boson~$A'$ interacting with charged particles via kinetic mixing with the photon. In these experiments, elastic scattering of electrons or positrons on a hydrogen gas target is used, and the process $e^{\pm} p \rightarrow e^{\pm} p \, \gamma$ is one of the main sources of background events.

The remainder of this paper is organized as follows. Section~\ref{S_2} contains the theoretical foundations needed to describe the elastic scattering of charged leptons on protons and the first-order QED radiative corrections to this process. In section~\ref{S_3} we discuss the practical aspects of using the generator to account for the standard radiative corrections in TPE measurements. Section~\ref{S_4} describes the basic algorithm that is used in \texttt{ESEPP} to generate events. Section~\ref{S_5} provides a summary. Finally, some technical details of the calculations are given in appendices~\hyperlink{app1}{A} and~\hyperlink{app2}{B}.

\section{Theoretical foundations}
\label{S_2}
\setcounter{equation}{0}

The values of the proton electromagnetic form factors can be determined from the Born differential cross section ${\rm d} \sigma_{\text{Born}} / {\rm d} \Omega_{\ell}$ of unpolarized elastic lepton--proton scattering obtained at fixed four-momentum transfer, but with different lepton scattering angles and incident beam energies. This is the above-mentioned Rosenbluth method, based on the one-photon exchange approximation (see diagram~(a) in figure~\ref{Fig2}). It should be remembered that any cross section measurement inevitably involves the contributions of higher-order QED processes~--- the radiative corrections to the Born cross section. Taking into account the first-order QED radiative corrections, we can write the following schematic expression for the cross section of charged leptons scattering on protons
\begin{gather}
\sigma (\ell^{\pm} p) \propto |{\mathcal M}_{\text{Born}}|^2 + 2\re{\bigl[{\mathcal M}_{\text{Born}}^{\dagger} \bigl({\mathcal M}_{\text{vac}} + {\mathcal M}_{\text{vert}}^{\ell} + {\mathcal M}_{\text{vert}}^p\bigr)\bigr]} \nonumber \\
{} + 2\re{\bigl[{\mathcal M}_{\text{Born}}^{\dagger} \bigl({\mathcal M}_{\text{box}} + {\mathcal M}_{\text{xbox}}\bigr)\bigr]} + \bigl|{\mathcal M}_{\text{brems}}^{\text{li}} + {\mathcal M}_{\text{brems}}^{\text{lf}}\bigr|^2 + \bigl|{\mathcal M}_{\text{brems}}^{\text{pi}} + {\mathcal M}_{\text{brems}}^{\text{pf}}\bigr|^2 \nonumber \\
{} + 2\re{\bigl[\bigl({\mathcal M}_{\text{brems}}^{\text{li}} + {\mathcal M}_{\text{brems}}^{\text{lf}}\bigr)^{\dagger} \bigl({\mathcal M}_{\text{brems}}^{\text{pi}} + {\mathcal M}_{\text{brems}}^{\text{pf}}\bigr)\bigr]} + {\mathcal O}(\alpha^4), \label{eq.2.1}
\end{gather}
where ${\mathcal M}_{\text{Born}}$ is the Born amplitude; ${\mathcal M}_{\text{vac}}$ is the vacuum polarization amplitude; ${\mathcal M}_{\text{vert}}^{\ell}$ and ${\mathcal M}_{\text{vert}}^p$ are the amplitudes related to the lepton and proton vertex corrections; ${\mathcal M}_{\text{box}}$ and ${\mathcal M}_{\text{xbox}}$ are the box and crossed-box TPE amplitudes; ${\mathcal M}_{\text{brems}}^{\text{li}}$, ${\mathcal M}_{\text{brems}}^{\text{lf}}$, ${\mathcal M}_{\text{brems}}^{\text{pi}}$, and ${\mathcal M}_{\text{brems}}^{\text{pf}}$ are the first-order bremsstrahlung amplitudes in the cases when a photon is emitted by the initial-state lepton, final-state lepton, initial-state proton, and final-state proton, respectively (see figure~\ref{Fig2}). The interference TPE term (the first term in the second line of the above expression) and the interference bremsstrahlung term (the third line) are charge-odd, i.e., they change sign depending on the sign of the lepton's charge. All other terms in~(\ref{eq.2.1}) are charge-even.

It is important to understand that all the above-mentioned amplitudes, except ${\mathcal M}_{\text{Born}}$ and ${\mathcal M}_{\text{vac}}$, contain infrared-divergent terms (tending to infinity in the limit of very soft, or `infrared', photons). These are canceled out completely in the sum~(\ref{eq.2.1}), which is therefore finite. More precisely, there are the following cancellations of the infrared divergences (see section~\ref{Ss_2.5} for details): between the lepton vertex correction and the lepton bremsstrahlung correction; between the proton vertex correction and the proton bremsstrahlung correction; between the TPE correction and the correction due to interference of the lepton and proton bremsstrahlung. 

It is convenient to separate the amplitude ${\mathcal M}_{2\gamma} = {\mathcal M}_{\text{box}} + {\mathcal M}_{\text{xbox}}$ into `soft' and `hard' parts, such that ${\mathcal M}_{2\gamma} = {\mathcal M}_{2\gamma}^{\text{soft}} + {\mathcal M}_{2\gamma}^{\text{hard}}$. The label `soft' implies the assumption that at least one virtual photon is soft. The soft part of this amplitude is independent of the proton structure, while the hard part is highly model-dependent. Infrared divergences are completely contained in the soft part, but it is important to remember that such a decomposition into soft and hard parts is not unique. The standard radiative corrections take into account only the soft part of the TPE amplitude, and the hard part is usually assumed to be negligible. The situation is complicated by the fact that the hard TPE corrections are difficult to calculate due to the presence of excited intermediate proton states. This is the reason why the hard TPE effects are investigated in the dedicated experiments~\cite{nucl-ex.0408020, NPBPS.225-227.216, AIPConfProc.1160.19, NIMA.741.1, AIPConfProc.1160.24, PRC.88.025210} (see section~\ref{S_3}).

The processes discussed above can be taken into account by introducing a correction~$\delta$ to the Born cross section, defined as
\begin{equation}
\frac{{\rm d} \sigma_{\text{meas}}}{{\rm d} \Omega_{\ell}} = (1 + \delta) \, \frac{{\rm d} \sigma_{\text{Born}}}{{\rm d} \Omega_{\ell}}, \label{eq.2.2}
\end{equation}
where ${\rm d} \sigma_{\text{meas}} / {\rm d} \Omega_{\ell}$ is the experimentally measured differential cross section. The quantity~$\delta$ includes both the corrections due to the emission of a real bremsstrahlung photon and the virtual-photon corrections (see figure~\ref{Fig2}). However, corrections of both these types are infrared-divergent, so that only the total correction~$\delta$ can be defined uniquely. The only exception is the vacuum polarization correction~$\delta_{\text{vac}}$, which is finite and therefore can be determined individually.

As it has been mentioned already, the size of the radiative corrections depends not only on the kinematics of elastic scattering (defined, for example, by the beam energy~$E_{\ell}$ and the lepton scattering angle~$\theta_{\ell}$), but also on the certain experimental conditions and cuts used to select elastic scattering events. Therefore, a realistic Monte Carlo simulation is required to carefully take into account the radiative corrections in the general case. However, the situation is simpler in the particular case of single-arm (inclusive) experiments, when only the scattered lepton is detected, typically using a magnetic spectrometer. In such a case, the procedure to select elastic scattering events can be described by the single parameter~$\Delta E$, which is the maximum allowable energy loss of the scattered lepton due to inelastic processes. This parameter means that the energy of the lepton, detected at the certain angle~$\theta_{\ell}$, should be in the range from $E_{\ell}^{\text{elas}} - \Delta E$ to $E_{\ell}^{\text{elas}}$, where $E_{\ell}^{\text{elas}} \approx M E_{\ell} / \left[M + E_{\ell} (1 - \cos{\theta_{\ell}})\right]$ is the elastic peak energy and $M$~is the proton mass.

Even in this simple case, we should remember that the angular and energy resolutions of any realistic detector are limited. The situation is much more complicated in the case of coincidence (exclusive) experiments, when both the scattered lepton and recoil proton are detected. It is then impossible to describe all the experimental conditions and cuts by a single parameter, making a careful Monte Carlo simulation necessary.

Let us mention some notable publications devoted to the radiative corrections to elastic electron--proton scattering. Mo and Tsai provided in their papers~\cite{PR.122.1898, RMP.41.205} the standard prescription for radiative corrections used for several decades in most single-arm experiments. Relatively recently, Maximon and Tjon published a paper~\cite{PRC.62.054320} revising this approach and removing many of the mathematical and physical approximations used by Mo and Tsai. The work~\cite{PRC.64.054610} adapts the formulas obtained by Mo and Tsai to the case of coincidence experiments and discusses the impact of higher-order bremsstrahlung effects. Note that there are several other approaches to the radiative corrections for $e^- p$~scattering (see~\cite{PR.130.1210, PRC.62.025501, PRD.64.113009, PRC.75.015207, arXiv:1311.0370}, for example), but their discussion is beyond the scope of this paper.

\begin{table}[t]
\centering
\begin{tabular}{|Sc|Sc|Sc|Sc|Sc|Sc|Sc|}
\hline Calculation & $1 + \delta^+$ & {\large $e^{\delta^+}$} & $1 + \delta^-$ & {\large $e^{\delta^-}$} & {\large $\frac{1 + \delta^+}{1 + \delta^-}$} & {\large $e^{\delta^+ - \delta^-}$} \\
\hline \hline Mo and Tsai & $0.9993$ & $0.9993$ & $0.9794$ & $0.9796$ & $1.0203$ & $1.0201$ \\
Maximon and Tjon & $1.0001$ & $1.0001$ & $0.9725$ & $0.9729$ & $1.0283$ & $1.0279$ \\
\hline
\end{tabular}
\caption{Values of the radiative corrections for the illustrative example with $E_{\ell} = 1~\text{GeV}$, $-q^2 = 1~\text{GeV}^2$, and $\Delta E = 0.1~\text{GeV}$.}
\label{Tab1}
\end{table}

Since we restrict ourselves to considering bremsstrahlung of the first order only, let us briefly examine the accuracy of this approximation. As shown by Yennie, Frautschi, and Suura~\cite{AnnPhys.13.379}, the emission of soft bremsstrahlung photons can be summed up to all orders in~$\alpha$ via exponentiation of~$\delta$
\begin{equation}
\frac{{\rm d} \sigma_{\text{meas}}}{{\rm d} \Omega_{\ell}} = \exp{(\delta)} \, \frac{{\rm d} \sigma_{\text{Born}}}{{\rm d} \Omega_{\ell}}. \label{eq.2.3}
\end{equation}
This exponentiation procedure is incompatible with our approach, but we can use the formula (\ref{eq.2.3}) to make a rough estimation of the contribution of higher-order bremsstrahlung. To do this, we choose the following numerical parameters approximately corresponding to the Novosibirsk TPE experiment: $E_{\ell} = 1~\text{GeV}$, $-q^2 = 1~\text{GeV}^2$, and $\Delta E = 0.1~\text{GeV}$. Further, we use the formula~(II.6) by Mo and Tsai~\cite{RMP.41.205} and the formula~(5.2) by Maximon and Tjon~\cite{PRC.62.054320} to calculate the radiative corrections $\delta^+$ and $\delta^-$ for the cases of $e^+ p$ and $e^- p$ scattering. The obtained numerical values are shown in table~\ref{Tab1}. Note that in the formula~(5.2) of Maximon and Tjon we have neglected the term~$\delta_{el}^{(1)}$ related to the model-dependent part of the amplitude~${\mathcal M}_{\text{vert}}^p$. As can be seen from table~\ref{Tab1}, in the case of $e^- p$ scattering using the formula~(\ref{eq.2.2}) instead of~(\ref{eq.2.3}) leads to a relative error in the extracted Born cross section of $0.02\%$ according to Mo and Tsai and $0.04\%$ according to Maximon and Tjon. These errors are even smaller in the case of $e^+ p$ scattering. It is important to note that using the calculation of Mo and Tsai instead of the more accurate calculation of Maximon and Tjon leads to a relative error of about $0.7\%$ in the cross section ${\rm d} \sigma_{\text{Born}} / {\rm d} \Omega_{\ell}$ for our example of $e^- p$ scattering. This error is much more significant than the error caused by neglecting higher-order bremsstrahlung. Finally, if we are interested in the ratio $R = \sigma (e^+ p) / \sigma (e^- p)$ of the $e^+ p$ and $e^- p$ elastic scattering cross sections, then the replacement of the radiative correction factor $(1 + \delta^+) / (1 + \delta^-)$ with the exponentiated one changes the extracted value of~$R$ by $0.02\%$ and~$0.04\%$ according to~\cite{RMP.41.205} and~\cite{PRC.62.054320}, respectively. Such corrections are negligible for the Novosibirsk TPE experiment.

Nevertheless, accounting for higher-order bremsstrahlung may be significant for some precise measurements. In particular, it becomes more important with decreasing~$\Delta E$. It is possible, but not straightforward, to further improve the described approach by taking into account the second-order bremsstrahlung process $\ell^{\pm} p \rightarrow \ell^{\pm} p \, \gamma \, \gamma$ (see~\cite{PRC.62.025501} also).

In the end of this introductory part, let us note that we use the system of units in which $\hbar = c = 1$~and $\alpha = e^2 / 4\pi \approx 1/137.036$. With this choice of units, all energies, momenta, and masses of elementary particles are expressed in~$\text{GeV}$ and scattering cross sections in $\text{GeV}^{-2}$ ($1~\text{GeV}^{-2} \approx 389~\text{microbarn}$). We use the metric such that $p_i \cdot p_j = E_i E_j - {\mathbf p}_i \cdot {\mathbf p}_j$. We consider only the laboratory frame, where the target proton is at rest. As usual, Feynman slash notation~$\slashed p$ represents $p_{\mu} \gamma^{\mu}$. For convenience, let us give here a list of some basic notations which we use throughout the paper:

\vspace{2mm}
\hspace{-6mm}
\begin{tabular}{rl}
$q^2$ & \hspace{-2mm}---~four-momentum transfer squared in the case of purely elastic scattering;\\
$m$, $M$ & \hspace{-2mm}---~masses of the charged lepton (electron or muon) and the proton;\\
$\mu$ \! & \hspace{-2mm}---~magnetic moment of the proton ($\mu \approx 2.7928$ in nuclear magnetons);\\
$G_E$, $G_M$ & \hspace{-2mm}---~electric and magnetic Sachs form factors of the proton;\\
$F_1$, $F_2$ & \hspace{-2mm}---~Dirac and Pauli form factors of the proton;\\
$E_{\ell}$, $E_{\ell}'$ & \hspace{-2mm}---~full energies of the incident and scattered lepton;\\
$E_p$ & \hspace{-2mm}---~full energy of the recoil proton;\\
$E_{\gamma}$ & \hspace{-2mm}---~energy of the bremsstrahlung photon;\\
$\theta_{\ell}$, $\varphi_{\ell}$ & \hspace{-2mm}---~polar and azimuthal angles of the scattered lepton;\\
$\theta_p$, $\varphi_p$ & \hspace{-2mm}---~polar and azimuthal angles of the recoil proton;\\
$\theta_{\gamma}$, $\varphi_{\gamma}$ & \hspace{-2mm}---~polar and azimuthal angles of the bremsstrahlung photon;\\
$\ell$, ${\boldsymbol \ell}$ \! & \hspace{-2mm}---~four- and three-momenta of the incident lepton ($|\boldsymbol \ell|^2 = E_{\ell}^2 - m^2$);\\
$\ell'$, ${\boldsymbol \ell}'$ & \hspace{-2mm}---~four- and three-momenta of the scattered lepton ($|{\boldsymbol \ell}'|^2 = {E_{\ell}'}^2 - m^2$);\\
$p$ \! & \hspace{-2mm}---~four-momentum of the target proton;\\
$p'$, ${\mathbf p}'$ & \hspace{-2mm}---~four- and three-momenta of the recoil proton ($|{\mathbf p}'|^2 = E_p^2 - M^2$);\\
$k$, ${\mathbf k}$ & \hspace{-2mm}---~four- and three-momenta of the bremsstrahlung photon.
\end{tabular}

\subsection{Elastic scattering of charged leptons on protons}
\label{Ss_2.1}

In the case of one-photon exchange (or in the first Born approximation) and assuming that $-q^2 \gg m^2$, the differential cross section for unpolarized elastic scattering of charged leptons on protons is given by the well-known Rosenbluth formula~\cite{PR.79.615}
\begin{equation}
\frac{{\rm d} \sigma_{\text{Born}}}{{\rm d} \Omega_{\ell}} = \frac{1}{\varepsilon (1 + \tau)} \left[\varepsilon \, G_E^2 (q^2) + \tau \, G_M^2 (q^2)\right] \frac{{\rm d} \sigma_{\text{Mott}}}{{\rm d} \Omega_{\ell}}, \label{eq.2.4}
\end{equation}
where $\tau = -q^2 / (4 M^2)$; $\varepsilon = \bigl[1 + 2 (1 + \tau) \tan^2{(\theta_{\ell} / 2)}\bigr]^{-1}$ is a convenient dimensionless variable, lying in the range $0 < \varepsilon < 1$ and describing the separation between the longitudinal (charge) and transverse (magnetic) parts of the cross section; ${\rm d} \sigma_{\text{Mott}} / {\rm d} \Omega_{\ell}$ is the Mott differential cross section including the proton recoil. Let us recall that the Mott cross section describes the scattering of charged leptons on spinless point charged particles and is expressed as
\begin{equation}
\frac{{\rm d} \sigma_{\text{Mott}}}{{\rm d} \Omega_{\ell}} = \frac{\alpha^2}{4 E_{\ell}^2} \, \frac{\cos^2{(\theta_{\ell} / 2)}}{\sin^4{(\theta_{\ell} / 2)}} \, \frac{E_{\ell}'}{E_{\ell}}. \label{eq.2.5}
\end{equation}

It is important to mention that in the general case, when the lepton mass~$m$ is not negligible, the Rosenbluth formula (\ref{eq.2.4}) is still valid provided that one uses the following expressions for~$\varepsilon$ and ${\rm d} \sigma_{\text{Mott}} / {\rm d} \Omega_{\ell}$ (see~\cite{PRC.36.2466}, for example):
\begin{gather}
\widetilde \varepsilon = \left[1 - 2 (1 + \tau) \, \frac{q^2 + 2 m^2}{4 E_{\ell} E_{\ell}' + q^2}\right]^{-1}, \label{eq.2.6} \\
\frac{{\rm d} \sigma_{\text{Mott}}}{{\rm d} \Omega_{\ell}} = \frac{\alpha^2}{4 E_{\ell}^2} \, \frac{1 + q^2 / (4 E_{\ell} E_{\ell}')}{q^4 / (4 E_{\ell} E_{\ell}')^2} \, \frac{1}{d} \, \frac{M ({E_{\ell}'}^2 - m^2)}{M E_{\ell} E_{\ell}' + m^2 (E_{\ell}' - E_{\ell} - M)}, \label{eq.2.7}
\end{gather}
where
\begin{equation}
q^2 = 2M (E_{\ell}' - E_{\ell}), \qquad d = \frac{E_{\ell}'}{E_{\ell}} \, \frac{|{\boldsymbol \ell}|}{|{\boldsymbol \ell}'|}. \label{eq.2.8}
\end{equation}
Note that the values of~$\,\widetilde \varepsilon\,$ larger than unity are possible (which is why we have marked this new variable with a tilde). For example, it is easily seen that in the limit $-q^2 \rightarrow 0$ the formula (\ref{eq.2.6}) gives $\widetilde \varepsilon = E_{\ell}^2 / (E_{\ell}^2 - m^2) > 1$. If the lepton mass~$m$ is negligible, then the new variable~$\widetilde \varepsilon$ coincides with the usual~$\varepsilon$.

To generate events according to the Rosenbluth formula~(\ref{eq.2.4}) we need to use a certain parametrization of the proton form factors. The simplest possibility is to use the dipole parametrization
\begin{equation}
G_E (q^2) = G_D (q^2), \qquad G_M (q^2) = \mu \, G_D (q^2), \qquad G_D (q^2) = \left(1 - \frac{q^2}{\Lambda^2}\right)^{\!-2}, \label{eq.2.9}
\end{equation}
where $\Lambda^2 = 0.71~\text{GeV}^2$. In particular, in the static limit with $q^2 = 0$ we have $G_E (0) = 1$ and $G_M (0) = \mu$. We will also use the Dirac and Pauli form factors, $F_1$ and $F_2$, which are expressed in terms of the Sachs form factors, $G_E$ and $G_M$, as
\begin{equation}
F_1 (q^2) = \frac{G_E (q^2) + \tau \, G_M (q^2)}{1 + \tau}, \qquad F_2 (q^2) = \frac{G_M (q^2) - G_E (q^2)}{1 + \tau}. \label{eq.2.10}
\end{equation}

Kelly proposed~\cite{PRC.70.068202} a more general parametrization of the proton form factors which is given by the ratio of a polynomial of order~$n$ to a polynomial of order~$n + 2$ in~$\tau$:
\begin{equation}
G_E (q^2) = G_1 (q^2), \qquad G_M (q^2) = \mu \, G_2 (q^2), \qquad G_i (q^2) = \frac{\sum_{k = 0}^n a_{ik} \tau^k}{1 + \sum_{k = 1}^{n + 2} b_{ik} \tau^k}, \label{eq.2.11}
\end{equation}
where $i = 1, 2$ and we should fix the coefficients $a_{i0} = 1$ to get the correct values for $G_E (0)$ and $G_M (0)$. For high values of the four-momentum transfer, the parametrization~(\ref{eq.2.11}) has the behavior $G_i \propto q^{-4}$ expected from dimensional scaling laws in perturbative quantum chromodynamics.

In the simplest possible case of~$n = 0$, the parametrization~(\ref{eq.2.11}) is exactly the same as the dipole parametrization~(\ref{eq.2.9}) after choosing $b_{i1} = 8M^2 / \Lambda^2$ and $b_{i2} = 16 M^4/ \Lambda^4$. In the next simplest case of~$n = 1$, $G_1 (q^2)$ and~$G_2 (q^2)$ from~(\ref{eq.2.11}) become
\begin{equation}
G_1 (q^2) = \frac{1 + a_{11} \tau}{1 + b_{11} \tau + b_{12} \tau^2 + b_{13} \tau^3}, \qquad G_2 (q^2) = \frac{1 + a_{21} \tau}{1 + b_{21} \tau + b_{22} \tau^2 + b_{23} \tau^3}. \label{eq.2.12}
\end{equation}

\begin{table}[t]
\centering
\begin{tabular}{|Sc|Sc|Sc|Sc|Sc|Sc|Sc|Sc|Sc|}
\hline Parametri- & \multicolumn{4}{Sc|}{$G_E$} & \multicolumn{4}{Sc|}{$G_M / \mu$} \\
\cline{2-9} zation & $a_{11}$ & $b_{11}$ & $b_{12}$ & $b_{13}$ & $a_{21}$ & $b_{21}$ & $b_{22}$ & $b_{23}$ \\
\hline \hline Dipole & $0$ & $9.92$ & $24.6$ & $0$ & $0$ & $9.92$ & $24.6$ & $0$ \\
Kelly~\cite{PRC.70.068202} & $-0.24$ & $10.98$ & $12.82$ & $21.97$ & $0.12$ & $10.97$ & $18.86$ & $6.55$ \\
Puckett~\cite{arXiv:1008.0855} & $-0.299$ & $11.11$ & $14.11$ & $15.7$ & $0.081$ & $11.15$ & $18.45$ & $5.31$ \\
\hline
\end{tabular}
\caption{Coefficients $a_{i1}$ and $b_{ik}$ for the discussed form factor parametrizations.}
\label{Tab2}
\end{table}

It was shown in the original paper~\cite{PRC.70.068202} that the parametrization~(\ref{eq.2.12}) provides reasonably good fits to the existing data for the proton form factors including those obtained in the polarization transfer measurements. Using the same parametrization, Puckett recently carried out a global fit~\cite{arXiv:1008.0855} taking into account the data of the GEp-III experiment~\cite{PRL.104.242301} at JLab. Numerical values of the coefficients~$a_{i1}$, $b_{i1}$, $b_{i2}$, and~$b_{i3}$ obtained by Kelly and Puckett are given in table~\ref{Tab2}. Figure~\ref{Fig1} shows that both of these fits are good at describing the polarization transfer data for~$\mu \, G_E / G_M$.

\subsection{Kinematics of the process\texorpdfstring{ ${\ell}^{\pm} p \rightarrow {\ell}^{\pm} p \, \gamma$}{}}
\label{Ss_2.2}

In this section we consider the kinematics of lepton--proton scattering accompanied by the emission of a single bremsstrahlung photon. Conservation of the four-momentum in this process gives
\begin{equation}
\ell + p = \ell' + p' + k. \label{eq.2.13}
\end{equation}
We choose our coordinate system such that the $z$~axis is directed along the momentum of the incident lepton and the lepton scatters in the $xz$~plane ($\varphi_{\ell} \equiv 0$). Then the four-momenta of the particles involved in the process can be written as follows:
\begin{gather}
\ell = \left(E_{\ell}, \, {\boldsymbol \ell}\right) = \left(E_{\ell}, \, 0, \, 0, \, |{\boldsymbol \ell}|\right), \label{eq.2.14} \\
p = \left(M, \, 0\right) = \left(M, \, 0, \, 0, \, 0\right), \label{eq.2.15} \\
\ell' = \left(E_{\ell}', \, {\boldsymbol \ell}'\right) = \left(E_{\ell}', \, |{\boldsymbol \ell}'| \sin{\theta_{\ell}}, \, 0, \, |{\boldsymbol \ell}'| \cos{\theta_{\ell}}\right), \label{eq.2.16} \\
p' = \left(E_p, \, {\mathbf p}'\right) = \left(E_p, \, |{\mathbf p}'| \sin{\theta_p} \cos{\varphi_p}, \, |{\mathbf p}'| \sin{\theta_p} \sin{\varphi_p}, \, |{\mathbf p}'| \cos{\theta_p}\right), \label{eq.2.17} \\
k = \left(E_{\gamma}, \, {\mathbf k}\right) = \left(E_{\gamma}, \, E_{\gamma} \sin{\theta_{\gamma}} \cos{\varphi_{\gamma}}, \, E_{\gamma} \sin{\theta_{\gamma}} \sin{\varphi_{\gamma}}, \, E_{\gamma} \cos{\theta_{\gamma}}\right). \label{eq.2.18}
\end{gather}

The full energy of the incident lepton, $E_{\ell}$, is assumed to be known. Then, all the kinematic parameters of the final-state particles can be expressed in terms of the variables $\theta_{\ell}$, $E_{\gamma}$, $\theta_{\gamma}$, and~$\varphi_{\gamma}$. In particular, the identity $(\ell + p - \ell' - k)^2 = (p')^2 = M^2$ leads to the following equation for the variable~$E_{\ell}'$:
\begin{equation}
A \sqrt{{E_{\ell}'}^2 - m^2} = B E_{\ell}' + C, \label{eq.2.19}
\end{equation}
where the coefficients $A$, $B$, and $C$ are
\begin{gather}
A = |{\boldsymbol \ell}| \cos{\theta_{\ell}} - E_{\gamma} \cos{\psi}, \label{eq.2.20} \\
\cos{\psi} = \cos{\theta_{\ell}} \cos{\theta_{\gamma}} + \sin{\theta_{\ell}} \sin{\theta_{\gamma}} \cos{\varphi_{\gamma}}, \label{eq.2.21} \\
B = E_{\ell} + M - E_{\gamma}, \label{eq.2.22} \\
C = E_{\gamma} \left(E_{\ell} + M - |{\boldsymbol \ell}| \cos{\theta_{\gamma}}\right) - M E_{\ell} - m^2. \label{eq.2.23}
\end{gather}

In order to find a solution to~(\ref{eq.2.19}), we need to square both sides of this expression. The two roots of the resulting quadratic equation are given by the formula
\begin{equation}
E_{\ell}' = \frac{B C \pm A \sqrt{m^2 (A^2 - B^2) + C^2}}{A^2 - B^2}. \label{eq.2.24}
\end{equation}
In most cases, only the sign `$-$' in~(\ref{eq.2.24}) provides a physical solution. However, in some kinematic region both roots~(\ref{eq.2.24}) can be physical. It should always be checked that the calculated values of~$E_{\ell}'$ satisfy the initial equation (\ref{eq.2.19}). The two other criteria of a physical solution that must be satisfied are
\begin{equation}
E_{\gamma} < \frac{M (E_{\ell} - m)}{M + E_{\ell} - |{\boldsymbol \ell}| \cos{\theta_{\gamma}}}, \qquad m < E_{\ell}' < E_{\ell} - E_{\gamma}. \label{eq.2.25}
\end{equation}

If the lepton mass can be neglected ($m \ll E_{\ell}, E_{\ell}'$), the value of the scattered lepton energy is unique and given by the simple formula
\begin{equation}
E_{\ell}' = \frac{C}{A - B} = \frac{M \left(E_{\ell} - E_{\gamma}\right) - E_{\ell} E_{\gamma} \left(1 - \cos{\theta_{\gamma}}\right)}{M + E_{\ell} \left(1 - \cos{\theta_{\ell}}\right) - E_{\gamma} \left(1 - \cos{\psi}\right)}. \label{eq.2.26}
\end{equation}
Another important special case is purely elastic scattering, when there is no bremsstrahlung photon. In this case, the value of~$E_{\ell}'$ is also defined uniquely
\begin{equation}
E_{\ell}' = \frac{\left(E_{\ell} + M\right) \left(M E_{\ell} + m^2\right) + \sqrt{M^2 - m^2 \sin^2{\theta_{\ell}}} \, |{\boldsymbol \ell}|^2 \cos{\theta_{\ell}}}{\left(E_{\ell} + M\right)^2 - |{\boldsymbol \ell}|^2 \cos^2{\theta_{\ell}}}. \label{eq.2.27}
\end{equation}

As soon as the value of~$E_{\ell}'$ is found, we know all the components of the four-momenta $\ell$, $p$, $\ell'$, and $k$. Then, the four-momentum~$p'$ can be easily calculated as $p' = \ell + p - \ell' - k$.

\subsection{Differential cross section of the process\texorpdfstring{ ${\ell}^{\pm} p \rightarrow {\ell}^{\pm} p \, \gamma$}{} in the soft-photon approximation}
\label{Ss_2.3}

In the soft-photon approximation, which is valid when $E_{\gamma} \ll E_{\ell}, E_{\ell}'$, the differential cross section of the process ${\ell}^{\pm} p \rightarrow {\ell}^{\pm} p \, \gamma$ is expressed through the Born cross section ${\rm d} \sigma_{\text{Born}} / {\rm d} \Omega_{\ell}$ by the following simple formula (see~\cite{PR.122.1898, PRC.64.054610}, for example):
\begin{equation}
\frac{{\rm d} \sigma_{\text{brems}}}{{\rm d} E_{\gamma} \, {\rm d} \Omega_{\gamma} \, {\rm d} \Omega_{\ell}} = E_{\gamma}^2 \, \frac{{\rm d} \sigma_{\text{brems}}}{{\rm d} \Omega_{\ell} \, {\rm d}^3 k} = -\frac{\alpha E_{\gamma}}{4\pi^2} \left[z \, \frac{\ell}{k \cdot \ell} - z \, \frac{\ell'}{k \cdot \ell'} + \frac{p}{k \cdot p} - \frac{p'}{k \cdot p'}\right]^2 \frac{{\rm d} \sigma_{\text{Born}}}{{\rm d} \Omega_{\ell}}, \label{eq.2.28}
\end{equation}
where one should choose the sign~$z = -1$ for the case of negatively charged leptons ($e^-$~or~$\mu^-$) and the sign~$z = +1$ for the case of positively charged leptons ($e^+$ or $\mu^+$). Using this formula, we can perform calculations in several different ways. 

In the simplest case, we can assume that emission of a soft photon does not violate the elastic kinematics of the scattering process. This leads to two consequences: first, we assume that the factor ${\rm d} \sigma_{\text{Born}} / {\rm d} \Omega_{\ell}$ in~(\ref{eq.2.28}) is the purely elastic cross section determined only by the incident lepton energy~$E_{\ell}$ and the scattering angle~$\theta_{\ell}$; secondly, we use the actual value of the photon four-momentum~$k$, but the elastic (unradiated) values of the fermion four-momenta $\ell'$ and~$p'$. We will refer to this approximation as the \emph{primary soft-photon approximation}. It is natural and its use greatly simplifies the analytical integration of the cross-section (\ref{eq.2.28}), as will be shown below. 

\begin{figure}[t]
\centering
\includegraphics[width=0.8\textwidth]{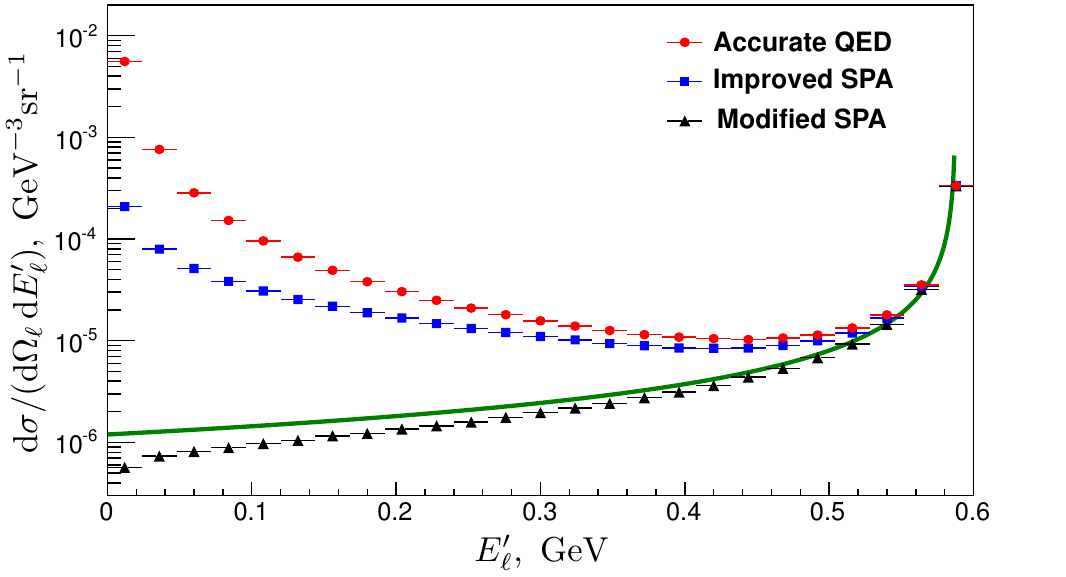}
\caption{Different predictions for the radiative tail. The kinematics are defined here by the incident electron energy, $E_{\ell} = 1~\text{GeV}$, and the scattering angle, $\theta_{\ell} = 70^{\circ}$. The elastic peak is located on the right side. The points are simulated using \texttt{ESEPP} with three different models for bremsstrahlung: the accurate QED calculation (red circles), as well as the improved (blue squares) and modified (black triangles) soft-photon approximations. The green curve represents a theoretical prediction according to~\cite{RMP.41.205, PRC.62.054320}.}
\label{Fig3}
\end{figure}

It is instructive, however, to consider other ways to perform calculations using the formula~(\ref{eq.2.28}). For example, we can use the actual (kinematically correct) values of the four-momenta $\ell'$ and $p'$ to calculate the factor in the square brackets. Let us call this approach the \emph{modified soft-photon approximation}. In addition to this, we can modify the cross section ${\rm d} \sigma_{\text{Born}} / {\rm d} \Omega_{\ell}$ if we calculate it using the usual formula (\ref{eq.2.4}), but assuming that the value of~$q^2$ is given now by~(\ref{eq.2.42}). This value of the four-momentum transfer corresponds to emission of a bremsstrahlung photon either by the incident or the scattered lepton, but not by the proton. It is a reasonable assumption, since the proton bremsstrahlung is usually suppressed in comparison with the lepton bremsstrahlung. We will refer to the approach we have described as the \emph{improved soft-photon approximation}. The adjective `improved' is rather arbitrary here, and we use it because this approach allows us to better reproduce the shape of the radiative tail in the case of relatively hard bremsstrahlung (see figure~\ref{Fig3}). Let us explain what we mean.

If the incident lepton loses energy due to emission of a hard photon then the probability for this lepton to be scattered by the proton increases. This phenomenon (noted in~\cite{RMP.41.205}, for example) can lead to the growth of the cross section with increasing energy of the emitted photon or, equivalently, to the rising of the radiative tail. The improved soft-photon approximation reproduces this effect to some extent, while the other two considered soft-photon approaches ignore it. A more accurate consideration of first-order bremsstrahlung beyond the soft-photon approximation is described in section~\ref{Ss_2.4}.

A quantitative comparison of the mentioned accurate calculation with the improved and modified soft-photon approximations can be found in figure~\ref{Fig3}, which shows different predictions for the cross section ${\rm d} \sigma / ({\rm d} \Omega_{\ell} \, {\rm d} E_{\ell}')$ as a function of the scattered electron energy for certain kinematics. The points in figure~\ref{Fig3} were obtained in a Monte Carlo simulation using \texttt{ESEPP}. A corresponding theoretical prediction can be made using a formula for $\delta (\Delta E)$ and the relationship between~$\Delta E$ and~$E_{\ell}'$. It is remarkable that the expressions~(II.6) from~\cite{RMP.41.205} and~(5.2) from~\cite{PRC.62.054320} give exactly the same shape for the radiative tail (see the green curve). Two conclusions can be drawn from figure~\ref{Fig3}. First, the improved soft-photon approximation is, indeed, closer to the accurate calculation, while the modified soft-photon approximation is closer to the analytical prediction. Second, all the predictions are in good agreement with each other in the vicinity of the elastic peak.

We also need to know the differential cross section of the process ${\ell}^{\pm} p \rightarrow {\ell}^{\pm} p \, \gamma$ in the case when the energy of the emitted photon is limited to a certain cut-off value~$E_{\gamma}^{\text{cut}}$. More precisely, we want to know the differential cross section ${\rm d} \sigma_{\text{brems}} / {\rm d} \Omega_{\ell}$ obtained from the cross section (\ref{eq.2.28}) by integration over all directions of the emitted photon and over its energy in the range $E_{\gamma} < E_{\gamma}^{\text{cut}}$. In the primary soft-photon approximation, the desired cross section is given by the expression
\begin{align}
\frac{{\rm d} \sigma_{\text{brems}}}{{\rm d} \Omega_{\ell}} \left.\rule{0mm}{6mm}\right|_{E_{\gamma} < E_{\gamma}^{\text{cut}}} &= \frac{-\alpha}{4\pi^2} \, \frac{{\rm d} \sigma_{\text{Born}}}{{\rm d} \Omega_{\ell}} \int\limits_{E_{\gamma} < E_{\gamma}^{\text{cut}}} \frac{{\rm d}^3 k}{E_\gamma} \left[z \, \frac{\ell}{k \cdot \ell} - z \, \frac{\ell'}{k \cdot \ell'} + \frac{p}{k \cdot p} - \frac{p'}{k \cdot p'}\right]^2 \nonumber \\
{} &= -2\alpha \, \frac{{\rm d} \sigma_{\text{Born}}}{{\rm d} \Omega_{\ell}} \sum_{i, j} \Theta (p_i) \Theta (p_j) B \bigl(p_i, p_j, E_{\gamma}^{\text{cut}}\bigr), \label{eq.2.29}
\end{align}
where
\begin{equation}
B \bigl(p_i, p_j, E_{\gamma}^{\text{cut}}\bigr) = \frac{1}{8\pi^2} \int\limits_{E_{\gamma} < E_{\gamma}^{\text{cut}}} \frac{{\rm d}^3 k}{\sqrt{|{\mathbf k}|^2 + \lambda^2}} \, \frac{p_i \cdot p_j}{(k \cdot p_i) (k \cdot p_j)}. \label{eq.2.30}
\end{equation}
Here, $i, j = 1, 2, 3, 4$ and the four-momenta~$p_1$, $p_2$, $p_3$, $p_4$ denote~$\ell$, $\ell'$, $p$, $p'$, respectively. The notation $\Theta (p_i)$ simply indicates the sign: $\Theta (\ell) = z$, $\Theta (\ell') = -z$, $\Theta (p) = 1$, and~$\Theta (p') = -1$. It has been assumed that the photon has a small fictitious mass~$\lambda$, which is necessary to make the integral~(\ref{eq.2.30}) convergent. The introduction of the parameter $\lambda$ is the standard method of regularization of the infrared divergences.

The calculation of~$B \bigl(p_i, p_j, E_{\gamma}^{\text{cut}}\bigr)$ is described in detail in appendix~\hyperlink{app1}{A}. We present here only the final result (note that similar expression in~\cite{PRC.64.054610} contains several misprints):
\begin{equation}
B \bigl(p_i, p_j, E_{\gamma}^{\text{cut}}\bigr) = \frac{p_i \cdot p_j}{4\pi} \int\limits_0^1 \frac{{\rm d} x}{p_x^2} \left(\ln{\frac{4 \bigl(E_{\gamma}^{\text{cut}}\bigr)^2}{p_x^2}} + \frac{p_x^0}{|{\mathbf p}_x|} \ln{\frac{p_x^0 - |{\mathbf p}_x|}{p_x^0 + |{\mathbf p}_x|}} + \ln{\frac{p_x^2}{\lambda^2}}\right), \label{eq.2.31}
\end{equation}
where we have introduced the four-momentum~$p_x$ defined as
\begin{equation}
p_x = \bigl(p_x^0, \, {\mathbf p}_x\bigr) = x p_i + (1 - x) p_j. \label{eq.2.32}
\end{equation}

The integral~(\ref{eq.2.31}) is divergent in the limit $\lambda \rightarrow 0$. Separating the divergent term, we can introduce the notation
\begin{gather}
\widetilde B \bigl(p_i, p_j, E_{\gamma}^{\text{cut}}\bigr) = B \bigl(p_i, p_j, E_{\gamma}^{\text{cut}}\bigr) - \frac{1}{4\pi} K (p_i, p_j) \nonumber \\
{} = \frac{p_i \cdot p_j}{4\pi} \int\limits_0^1 \frac{{\rm d} x}{p_x^2} \left(\ln{\frac{4 \bigl(E_{\gamma}^{\text{cut}}\bigr)^2}{p_x^2}} + \frac{p_x^0}{|{\mathbf p}_x|} \ln{\frac{p_x^0 - |{\mathbf p}_x|}{p_x^0 + |{\mathbf p}_x|}}\right), \label{eq.2.33}
\end{gather}
where
\begin{equation}
K (p_i, p_j) = (p_i \cdot p_j) \int\limits_0^1 \frac{{\rm d} x}{p_x^2} \, \ln{\frac{p_x^2}{\lambda^2}} \label{eq.2.34}
\end{equation}
and, in particular,
\begin{equation}
K (p_i, p_i) = \ln{\frac{m_i^2}{\lambda^2}}. \label{eq.2.35}
\end{equation}

The value of~$\widetilde B \bigl(p_i, p_j, E_{\gamma}^{\text{cut}}\bigr)$ is finite and can be calculated using the formula~(\ref{eq.2.33}), if the four-momenta $p_i$, $p_j$ and the cut-off energy~$E_{\gamma}^{\text{cut}}$ are known. Integration with respect to~$x$ can be easily carried out numerically. Note that in the case of $p_i = p_j$ this integration is trivial and can be done analytically
\begin{equation}
\widetilde B \bigl(p_i, p_i, E_{\gamma}^{\text{cut}}\bigr) = \frac{1}{2\pi} \left(\ln{\frac{2 E_{\gamma}^{\text{cut}}}{m_i}} + \frac{E_i}{\sqrt{E_i^2 - m_i^2}} \, \ln{\frac{m_i}{E_i + \sqrt{E_i^2 - m_i^2}}}\right). \label{eq.2.36}
\end{equation}
However, the formula~(\ref{eq.2.36}) does not apply directly to the case of~$p_i = p_j = p$, since the indeterminate form of type~$0/0$ appears. In this case, the correct expression, obtained by applying the l'H\^opital's rule to the second term in~(\ref{eq.2.36}), is
\begin{equation}
\widetilde B \bigl(p, p, E_{\gamma}^{\text{cut}}\bigr) = \frac{1}{2\pi} \left(\ln{\frac{2 E_{\gamma}^{\text{cut}}}{M}} - 1\right). \label{eq.2.37}
\end{equation}

For further convenience, let us write the expression~(\ref{eq.2.29}) explicitly
\begin{gather}
\frac{{\rm d} \sigma_{\text{brems}}}{{\rm d} \Omega_{\ell}} \left.\rule{0mm}{6mm}\right|_{E_{\gamma} < E_{\gamma}^{\text{cut}}} = -2\alpha \Bigl[\widetilde B \bigl(\ell, \ell, E_{\gamma}^{\text{cut}}\bigr) - 2\widetilde B \bigl(\ell, \ell', E_{\gamma}^{\text{cut}}\bigr) + \widetilde B \bigl(\ell', \ell', E_{\gamma}^{\text{cut}}\bigr) + 2z \widetilde B \bigl(\ell, p, E_{\gamma}^{\text{cut}}\bigr) \nonumber \\
{} - 2z \widetilde B \bigl(\ell, p', E_{\gamma}^{\text{cut}}\bigr) - 2z \widetilde B \bigl(\ell', p, E_{\gamma}^{\text{cut}}\bigr) + 2z \widetilde B \bigl(\ell', p', E_{\gamma}^{\text{cut}}\bigr) + \widetilde B \bigl(p, p, E_{\gamma}^{\text{cut}}\bigr) - 2\widetilde B \bigl(p, p', E_{\gamma}^{\text{cut}}\bigr) \nonumber \\
{} + \widetilde B \bigl(p', p', E_{\gamma}^{\text{cut}}\bigr) + \frac{1}{2\pi} \ln{\frac{m^2}{\lambda^2}} + \frac{1}{2\pi} \ln{\frac{M^2}{\lambda^2}} - \frac{1}{2\pi} K (\ell, \ell') + \frac{z}{2\pi} K (\ell, p) - \frac{z}{2\pi} K (\ell, p') \nonumber \\
{} - \frac{z}{2\pi} K (\ell', p) + \frac{z}{2\pi} K (\ell', p') - \frac{1}{2\pi} K (p, p')\Bigr] \, \frac{{\rm d} \sigma_{\text{Born}}}{{\rm d} \Omega_{\ell}}. \label{eq.2.38}
\end{gather}
We will show in section~\ref{Ss_2.5} that all the divergent terms in~(\ref{eq.2.38}) are canceled exactly with the corresponding terms arising from the virtual-photon radiative corrections.

\subsection{Differential cross section of the process\texorpdfstring{ ${\ell}^{\pm} p \rightarrow {\ell}^{\pm} p \, \gamma$}{} beyond the soft-photon and ultrarelativistic approximations}
\label{Ss_2.4}

The differential cross section of the process ${\ell}^{\pm} p \rightarrow {\ell}^{\pm} p \, \gamma$ is expressed in terms of its squared amplitude, $|{\mathcal M}_{\text{brems}}|^2$, as follows (see appendix~\hyperlink{app2}{B} for details):
\begin{equation}
\frac{{\rm d} \sigma_{\text{brems}}}{{\rm d} E_{\gamma} \, {\rm d} \Omega_{\gamma} \, {\rm d} \Omega_{\ell}} = \frac{1}{(4\pi)^5} \, \frac{1}{M |{\boldsymbol \ell}|} \sum_{E_{\ell}'} \frac{E_{\gamma} \, |{\boldsymbol \ell}'|^2 \, |{\mathcal M}_{\text{brems}}|^2}{\bigl|A E_{\ell}' - B |{\boldsymbol \ell}'|\bigr|}, \label{eq.2.39}
\end{equation}
where
\begin{equation}
|{\mathcal M}_{\text{brems}}|^2 = \bigl|{\mathcal M}_{\text{brems}}^{\ell}\bigr|^2 + \bigl|{\mathcal M}_{\text{brems}}^p\bigr|^2 + 2\re{\bigl({\mathcal M}_{\text{brems}}^{\ell \dagger} {\mathcal M}_{\text{brems}}^p\bigr)} \label{eq.2.40}
\end{equation}
and the summation should be performed over two values of~$E_{\ell}'$ in the case when both roots in~(\ref{eq.2.24}) are physical. The squared amplitude~$|{\mathcal M}_{\text{brems}}|^2$ is written in~(\ref{eq.2.40}) as the sum of three terms: the lepton term, $\bigl|{\mathcal M}_{\text{brems}}^{\ell}\bigr|^2 = \bigl|{\mathcal M}_{\text{brems}}^{\text{li}} + {\mathcal M}_{\text{brems}}^{\text{lf}}\bigr|^2$, corresponding to the case when the lepton emits a photon before or after scattering; the proton term, $\bigl|{\mathcal M}_{\text{brems}}^{p}\bigr|^2 = \bigl|{\mathcal M}_{\text{brems}}^{\text{pi}} + {\mathcal M}_{\text{brems}}^{\text{pf}}\bigr|^2$, corresponding to the case when the proton emits a photon before or after scattering; and the interference term, $2\re{\bigl({\mathcal M}_{\text{brems}}^{\ell \dagger} {\mathcal M}_{\text{brems}}^p\bigr)}$. Since we consider unpolarized scattering, these terms are averaged over the polarizations of the initial particles and summed over the polarizations of the final particles. Let us emphasize that the formula (\ref{eq.2.39}) is obtained without neglecting the lepton mass~$m$ (i.e., without using the ultrarelativistic approximation).

To calculate the lepton, proton, and interference terms in the framework of QED, we need to consider the Feynman diagrams (b)--(e) shown in figure~\ref{Fig2}. We assume that the intermediate hadronic states arising in the diagrams (d), (e) are the virtual-proton ones and that the vertices of the photon--proton interaction are described by the on-shell vertex operator~$\Gamma^{\mu}$ both for the virtual and real (bremsstrahlung) photons. It is known that in the general case the vertex of the interaction of an off-shell proton with a photon can be characterized by six invariant functions (see~\cite{PRC.42.599} and references therein). However, in the case when incoming (with four-momentum~$p_1$) and outgoing (with four-momentum~$p_2$) protons are on the mass shell (so that $p_1^2 = p_2^2 = M^2$), this vertex is described by the operator~$\Gamma^{\mu}$, defined as
\begin{equation}
\Gamma^{\mu} = F_1 (q^2) \, \gamma^{\mu} + \frac{F_2 (q^2)}{2 M} \, i \sigma^{\mu \nu} q_{\nu}, \label{eq.2.41}
\end{equation}
where $\sigma^{\mu \nu} = \frac{1}{2} i [\gamma^{\mu}, \gamma^{\nu}]$ and $q = p_2 - p_1$. Assuming the vertex operator (\ref{eq.2.41}), we can easily calculate the desired amplitudes without using the soft-photon and ultrarelativistic approximations. We refer to this calculation as the \emph{accurate QED calculation} in the sense that it is beyond the indicated approximations.

One should perform contractions of the lepton and proton tensors provided below to express the lepton, proton, and interference terms (\ref{eq.2.40}) through the scalar products $p_i \cdot p_j$ of the four-momenta $\ell$, $\ell'$, $p$, $p'$, and~$k$. We used the \texttt{FeynCalc} package~\cite{FeynCalc} designed for Mathematica to perform these contractions (details of the calculations can be found in~\cite{ESEPP}). For convenience, we introduce the following notation:
\begin{gather}
q_1^2 = \bigl(p' - p\bigr)^2 = 2 M \bigl(E_{\ell}' - E_{\ell} + E_{\gamma}\bigr), \label{eq.2.42} \\
q_2^2 = \bigl(\ell - \ell'\bigr)^2 = 2 |{\boldsymbol \ell}| |{\boldsymbol \ell}'| \cos{\theta_{\ell}} - 2 E_{\ell} E_{\ell}' + 2 m^2, \label{eq.2.43} \\
P = p + p', \qquad P_{+} = P + k, \qquad P_{-} = P - k, \label{eq.2.44}
\end{gather}
where $q_1^2$ and $q_2^2$ are the squares of the four-momenta transferred to the proton in the cases when the photon was emitted by the lepton and proton, respectively. Note that $q_1^2 = q_2^2 = q^2$ in the limit when $E_{\gamma} = 0$.

The lepton term is calculated as
\begin{equation}
\bigl|{\mathcal M}_{\text{brems}}^{\ell}\bigr|^2 = \frac{e^6}{q_1^4} \bigl({\mathcal L}_{1\mu \nu} + {\mathcal L}_{2\mu \nu} \bigr) {\mathcal P}^{\mu \nu}, \label{eq.2.45}
\end{equation}
where
\begin{gather}
{\mathcal L}_{1\mu \nu} = \frac{1}{2} \tr{\Bigl[\bigl(\slashed{\ell}' + m\bigr) \gamma^{\alpha} \frac{\slashed{\ell}' + \slashed{k} + m}{2 (k \cdot \ell')} \gamma_{\mu} \bigl(\slashed{\ell} + m\bigr) \gamma_{\alpha} \frac{\slashed{\ell} - \slashed{k} + m}{2 (k \cdot \ell)} \gamma_{\nu}\Bigr]} \nonumber \\
{} - \frac{1}{2} \tr{\Bigl[\bigl(\slashed{\ell}' + m\bigr) \gamma^{\alpha} \frac{\slashed{\ell}' + \slashed{k} + m}{2 (k \cdot \ell')} \gamma_{\mu} \bigl(\slashed{\ell} + m\bigr) \gamma_{\nu} \frac{\slashed{\ell}' + \slashed{k} + m}{2 (k \cdot \ell')} \gamma_{\alpha}\Bigr]}, \label{eq.2.46} \\
{\mathcal P}^{\mu \nu} = \frac{1}{2} \tr{\Bigl[\bigl(\slashed{p} + M\bigr)\Bigl\{\bigl(F_1 (q_1^2) + F_2 (q_1^2)\bigr) \gamma^{\nu} - \frac{F_2 (q_1^2)}{2 M} P^{\nu}\Bigr\}} \nonumber \\
\bigl(\slashed{p}' + M\bigr) \Bigr\{\bigl(F_1 (q_1^2) + F_2 (q_1^2)\bigr) \gamma^{\mu} - \frac{F_2 (q_1^2)}{2 M} P^{\mu}\Bigl\}\Bigr], \label{eq.2.47}
\end{gather}
and an expression for the lepton tensor~${\mathcal L}_{2\mu \nu}$ is obtained from the expression~(\ref{eq.2.46}) for~${\mathcal L}_{1\mu \nu}$ after changing~$\ell \leftrightarrow -\ell'$. The proton term is given by the formula
\begin{equation}
\bigl|{\mathcal M}_{\text{brems}}^p\bigr|^2 = \frac{e^6}{q_2^4} \, {\mathcal L}_{\mu \nu} \bigl({\mathcal P}_1^{\mu \nu} + {\mathcal P}_2^{\mu \nu}\bigr), \label{eq.2.48}
\end{equation}
where
\begin{gather}
{\mathcal L}_{\mu \nu} = \frac{1}{2} \tr{\Bigl[\bigl(\slashed{\ell} + m\bigr) \gamma_{\nu} \bigl(\slashed{\ell}' + m\bigr) \gamma_{\mu}\Bigr]}, \label{eq.2.49} \\
{\mathcal P}_1^{\mu \nu} = \frac{1}{2} \tr{\Bigl[\bigl(\slashed{p}' + M\bigr) \Bigl\{\bigl(F_1 (0) + F_2 (0)\bigr) \gamma^{\alpha}} \nonumber \\
{} - \frac{F_2 (0)}{2 M} \left[\bigl(2p' + k\bigr)^{\alpha} - \gamma^{\alpha} \bigl(\slashed{p}' + \slashed{k} - M\bigr)\right]\Bigr\} \frac{\slashed{p}' + \slashed{k} + M}{2 (k \cdot p')} \nonumber \\
\Bigl\{\bigl(F_1 (q_2^2) + F_2 (q_2^2)\bigr) \gamma^{\mu} - \frac{F_2 (q_2^2)}{2 M} \left[P_{+}^{\mu} - \bigl(\slashed{p}' + \slashed{k} - M\bigr) \gamma^{\mu}\right]\Bigr\} \bigl(\slashed{p} + M\bigr) \nonumber \\
\Bigl\{\bigl(F_1 (0) + F_2 (0)\bigr) \gamma^{\alpha} - \frac{F_2 (0)}{2 M} \left[\bigl(2p - k\bigr)^{\alpha} - \gamma^{\alpha} \bigl(\slashed{p} - \slashed{k} - M\bigr)\right]\Bigr\} \frac{\slashed{p} - \slashed{k} + M}{2 (k \cdot p)} \nonumber \\
\Bigl\{\bigl(F_1 (q_2^2) + F_2 (q_2^2)\bigr) \gamma^{\nu} - \frac{F_2 (q_2^2)}{2 M} \left[P_{-}^{\nu} - \bigl(\slashed{p} - \slashed{k} - M\bigr) \gamma^{\nu}\right]\Bigr\}\Bigr] \nonumber \\
{} - \frac{1}{2} \tr{\Bigl[\bigl(\slashed{p}' + M\bigr) \Bigl\{\bigl(F_1 (0) + F_2 (0)\bigr) \gamma^{\alpha}} \nonumber \\
{} - \frac{F_2 (0)}{2 M} \left[\bigl(2p' + k\bigr)^{\alpha} - \gamma^{\alpha} \bigl(\slashed{p}' + \slashed{k} - M\bigr)\right]\Bigr\} \frac{\slashed{p}' + \slashed{k} + M}{2 (k \cdot p')} \nonumber \\
\Bigl\{\bigl(F_1 (q_2^2) + F_2 (q_2^2)\bigr) \gamma^{\mu} - \frac{F_2 (q_2^2)}{2 M} \left[P_{+}^{\mu} - \bigl(\slashed{p}' + \slashed{k} - M\bigr) \gamma^{\mu}\right]\Bigr\} \bigl(\slashed{p} + M\bigr) \nonumber \\
\Bigl\{\bigl(F_1 (q_2^2) + F_2 (q_2^2)\bigr) \gamma^{\nu} - \frac{F_2 (q_2^2)}{2 M} \left[P_{+}^{\nu} - \gamma^{\nu} \bigl(\slashed{p}' + \slashed{k} - M\bigr)\right]\Bigr\} \frac{\slashed{p}' + \slashed{k} + M}{2 (k \cdot p')} \nonumber \\
\Bigl\{\bigl(F_1 (0) + F_2 (0)\bigr) \gamma^{\alpha} - \frac{F_2 (0)}{2 M} \left[\bigl(2p' + k\bigr)^{\alpha} - \bigl(\slashed{p}' + \slashed{k} - M\bigr) \gamma^{\alpha}\right]\Bigr\}\Bigr], \label{eq.2.50}
\end{gather}
and an expression for the proton tensor~${\mathcal P}_2^{\mu \nu}$ is obtained from the expression~(\ref{eq.2.50}) for~${\mathcal P}_1^{\mu \nu}$ after changing~$p \leftrightarrow -p'$.

Finally, the interference term can be calculated using the formula
\begin{equation}
2\re{\bigl({\mathcal M}_{\text{brems}}^{\ell \dagger} {\mathcal M}_{\text{brems}}^p\bigr)} = -z \frac{e^6}{q_1^2 q_2^2} \left(\frac{1}{k \cdot \ell'} \, {\mathcal L}_{1\mu \nu}^{\alpha} + \frac{1}{k \cdot \ell} \, {\mathcal L}_{2\mu \nu}^{\alpha} \right) \bigl({\mathcal P}_{1\alpha}^{\mu \nu} - {\mathcal P}_{2\alpha}^{\mu \nu}\bigr), \label{eq.2.51}
\end{equation}
where
\begin{gather}
{\mathcal L}_{1\mu \nu}^{\alpha} = \frac{1}{2} \tr{\Bigl[\bigl(\slashed{\ell}' + m\bigr) \gamma^{\alpha} \bigl(\slashed{\ell}' + \slashed{k} + m\bigr) \gamma_{\mu} \bigl(\slashed{\ell} + m\bigr) \gamma_{\nu}\Bigr]}, \label{eq.2.52} \\
{\mathcal P}_{1\alpha}^{\mu \nu} = \frac{1}{2} \tr{\Bigl[\bigl(\slashed{p}' + M\bigr) \Bigl\{\bigl(F_1 (q_1^2) + F_2 (q_1^2)\bigr) \gamma^{\mu} - \frac{F_2 (q_1^2)}{2 M} P^{\mu}\Bigr\} \bigl(\slashed{p} + M\bigr)} \nonumber \\
\Bigl\{\bigl(F_1 (q_2^2) + F_2 (q_2^2)\bigr) \gamma^{\nu} - \frac{F_2 (q_2^2)}{2 M} \left[P_{+}^{\nu} - \gamma^{\nu} \bigl(\slashed{p}' + \slashed{k} - M\bigr)\right]\Bigr\} \frac{\slashed{p}' + \slashed{k} + M}{2 (k \cdot p')} \nonumber \\
\Bigl\{\bigl(F_1 (0) + F_2 (0)\bigr) \gamma_{\alpha} - \frac{F_2 (0)}{2 M} \left[\bigl(2p' + k\bigr)_{\alpha} - \bigl(\slashed{p}' + \slashed{k} - M\bigr) \gamma_{\alpha}\right]\Bigr\}\Bigr], \label{eq.2.53}
\end{gather}
and expressions for the lepton~${\mathcal L}_{2\mu \nu}^{\alpha}$ and proton~${\mathcal P}_{2\alpha}^{\mu \nu}$ tensors are obtained from the expressions~(\ref{eq.2.52}) for~${\mathcal L}_{1\mu \nu}^{\alpha}$ and~(\ref{eq.2.53}) for~${\mathcal P}_{1\alpha}^{\mu \nu}$ after changing $\ell \leftrightarrow -\ell'$ and~$p \leftrightarrow -p'$, respectively.

\subsection{Virtual-photon radiative corrections and cancellation of the infrared divergences}
\label{Ss_2.5}

Let us start with the standard approach of Mo and Tsai~\cite{PR.122.1898, RMP.41.205} to the virtual-photon radiative corrections (using the on-shell renormalization scheme). Assuming $-q^2 \gg m^2$, all considered amplitudes are expressed through the Born amplitude ${\mathcal M}_{\text{Born}}$ as follows:
\begin{gather}
{\mathcal M}_{\text{vac}}^{e} = \frac{\alpha}{3\pi} \left[-\frac{5}{3} + \ln{\frac{-q^2}{m_e^2}}\right] {\mathcal M}_{\text{Born}}, \label{eq.2.54} \\
{\mathcal M}_{\text{vert}}^{\ell} = -\frac{\alpha}{2\pi} \left[K (\ell, \ell') - K (\ell, \ell) - \frac{3}{2} \ln{\frac{-q^2}{m^2}} + 2\right] {\mathcal M}_{\text{Born}}, \label{eq.2.55} \\
{\mathcal M}_{\text{vert}}^p = -\frac{\alpha}{2\pi} \bigl[K (p, p') - K (p, p)\bigr] {\mathcal M}_{\text{Born}}, \label{eq.2.56} \\
{\mathcal M}_{\text{box}}^{\text{MTs}} = z\frac{\alpha}{2\pi} \bigl[K (\ell, -p) + K (\ell', -p')\bigr] {\mathcal M}_{\text{Born}}, \label{eq.2.57} \\
{\mathcal M}_{\text{xbox}}^{\text{MTs}} = -z\frac{\alpha}{2\pi} \bigl[K (\ell', p) + K (\ell, p')\bigr] {\mathcal M}_{\text{Born}}, \label{eq.2.58}
\end{gather}
where the amplitude~${\mathcal M}_{\text{vac}}^e$ describes the contribution of electron--positron loops to the vacuum polarization and the infrared-divergent terms~$K (p_i, p_j)$ have the form~(\ref{eq.2.34}). The term $K (\ell, \ell)$ in~(\ref{eq.2.55}) was introduced by the renormalization of the amplitude~${\mathcal M}_{\text{vert}}^{\ell}$ and represents the infrared term of the electromagnetic lepton self-energy. Similarly, the term~$K (p, p)$ in~(\ref{eq.2.56}) represents the proton self-energy. 

The terms~$K (\ell, -p)$ and~$K (\ell', -p')$ in~(\ref{eq.2.57}) are complex, but only their real parts contribute to the cross section. For this reason, Mo and Tsai resorted to the following simplification of these terms:
\begin{equation}
\re{K (\ell, -p) \approx K (\ell, p)}, \qquad \re{K (\ell', -p') \approx K (\ell', p')}, \label{eq.2.59}
\end{equation}
which implies
\begin{equation}
{\mathcal M}_{\text{box}}^{\text{MTs}} \approx z\frac{\alpha}{2\pi} \bigl[K (\ell, p) + K (\ell', p')\bigr] {\mathcal M}_{\text{Born}}. \label{eq.2.60}
\end{equation}

The squared amplitude for purely elastic scattering, $|{\mathcal M}_{\text{elast}}|^2$, can be written as
\begin{equation}
|{\mathcal M}_{\text{elast}}|^2 = |{\mathcal M}_{\text{Born}}|^2 + \sum\limits_{i} 2\re{\bigl({\mathcal M}_{\text{Born}}^{\dag} {\mathcal M}_i\bigr)} + {\mathcal O} (\alpha^4), \label{eq.2.61}
\end{equation}
where the summation is over the amplitudes ${\mathcal M}_{\text{vac}}^e$, ${\mathcal M}_{\text{vert}}^{\ell}$, ${\mathcal M}_{\text{vert}}^p$, ${\mathcal M}_{\text{box}}^{\text{MTs}}$, and~${\mathcal M}_{\text{xbox}}^{\text{MTs}}$, so one obtains the following formula for the differential cross section~${\rm d} \sigma_{\text{elast}} / {\rm d} \Omega_{\ell}$:
\begin{gather}
\frac{{\rm d} \sigma_{\text{elast}}}{{\rm d} \Omega_{\ell}} = \Bigl\{1 + \frac{2\alpha}{3\pi} \Bigl[-\frac{5}{3} + \ln{\frac{-q^2}{m_e^2}}\Bigr] + \frac{\alpha}{\pi} \Bigl[\frac{3}{2} \ln{\frac{-q^2}{m^2}} - 2\Bigr] + \frac{\alpha}{\pi} \Bigl[-K (\ell, \ell') + \ln{\frac{m^2}{\lambda^2}} \nonumber \\
{} - K (p, p') + \ln{\frac{M^2}{\lambda^2}} + z K (\ell, p) + z K (\ell', p') - z K (\ell', p) - z K (\ell, p') \Bigr]\Bigr\} \, \frac{{\rm d} \sigma_{\text{Born}}}{{\rm d} \Omega_{\ell}}. \label{eq.2.62}
\end{gather}

The cross sections for both purely elastic scattering~(\ref{eq.2.62}) and scattering involving emission of a photon with the energy~$E_{\gamma} < E_{\gamma}^{\text{cut}}$~(\ref{eq.2.38}) contain infrared-divergent terms. However, the sum of these cross sections is finite, and only this sum has the physical meaning:
\begin{equation}
\frac{{\rm d} \sigma_{\text{elast}}}{{\rm d} \Omega_{\ell}} + \frac{{\rm d} \sigma_{\text{brems}}}{{\rm d} \Omega_{\ell}} \left.\rule{0mm}{6mm}\right|_{E_{\gamma} < E_{\gamma}^{\text{cut}}} = \bigl(1 + \delta_{\text{virt}} + \delta_{\text{brems}}\bigr) \, \frac{{\rm d} \sigma_{\text{Born}}}{{\rm d} \Omega_{\ell}}, \label{eq.2.63}
\end{equation}
where
\begin{gather}
\delta_{\text{virt}} = \delta_{\text{vac}}^e + \delta_{\text{vert}}, \label{eq.2.64} \\
\delta_{\text{brems}} = \delta_{\text{brems}}^{\ell \ell} + \delta_{\text{brems}}^{p p} + \delta_{\text{brems}}^{\ell p}, \label{eq.2.65}
\end{gather}
\begin{gather}
\delta_{\text{vac}}^{e} = \frac{2\alpha}{3\pi} \left(-\frac{5}{3} + \ln{\frac{-q^2}{m_e^2}}\right), \label{eq.2.66} \\
\delta_{\text{vert}} = \frac{\alpha}{\pi} \left(\frac{3}{2} \ln{\frac{-q^2}{m^2}} - 2\right), \label{eq.2.67} \\
\delta_{\text{brems}}^{\ell \ell} = -2\alpha \left[\widetilde B \bigl(\ell, \ell, E_{\gamma}^{\text{cut}}\bigr) - 2\widetilde B \bigl(\ell, \ell', E_{\gamma}^{\text{cut}}\bigr) + \widetilde B \bigl(\ell', \ell', E_{\gamma}^{\text{cut}}\bigr)\right], \label{eq.2.68} \\
\delta_{\text{brems}}^{pp} = -2\alpha \left[\widetilde B \bigl(p, p, E_{\gamma}^{\text{cut}}\bigr) - 2\widetilde B \bigl(p, p', E_{\gamma}^{\text{cut}}\bigr) + \widetilde B \bigl(p', p', E_{\gamma}^{\text{cut}}\bigr)\right], \label{eq.2.69} \\
\delta_{\text{brems}}^{\ell p} = -4 z \alpha \left[\widetilde B \bigl(\ell, p, E_{\gamma}^{\text{cut}}\bigr) - \widetilde B \bigl(\ell, p', E_{\gamma}^{\text{cut}}\bigr) - \widetilde B \bigl(\ell', p, E_{\gamma}^{\text{cut}}\bigr) + \widetilde B \bigl(\ell', p', E_{\gamma}^{\text{cut}}\bigr)\right]. \label{eq.2.70}
\end{gather}
Though the sum $\delta_{\text{virt}} + \delta_{\text{brems}}$ is defined uniquely, $\delta_{\text{virt}}$ and $\delta_{\text{brems}}$ individually are not because they are dependent on the procedure for divergence cancellation. The only exception is the correction $\delta_{\text{vac}}^{e}$, which is not infrared-divergent. For convenience, we have excluded the infrared-divergent terms in the formulas for $\delta_{\text{virt}}$ and $\delta_{\text{brems}}$, since all these terms finally disappear in~(\ref{eq.2.63}).

The expressions for the virtual-photon radiative corrections given above can be further refined. For example, the contributions of the electron, muon, and tau-lepton loops to the vacuum polarization are described by the general formula~\cite{Berestetskii}
\begin{equation}
\delta_{\text{vac}}^{e, \, \mu, \, \tau} = \frac{2\alpha}{3\pi} \left(-\frac{5}{3} + \frac{4\xi}{(1 - \xi)^2} - \frac{1 - 4\xi + \xi^2}{(1 - \xi)^2} \, \frac{1 + \xi}{1 - \xi} \ln{\xi}\right), \label{eq.2.71}
\end{equation}
where
\begin{equation}
\xi = -\frac{4 m_{\ell}^2}{q^2} \left(1 + \sqrt{1 - 4m_{\ell}^2 / q^2}\,\right)^{\!-2} \label{eq.2.72}
\end{equation}
and $m_{\ell}$ is the mass of the electron, muon or tau~lepton, respectively. In the approximation $-q^2 \gg m_{\ell}^2$, which is usually very good for~$m_{\ell} = m_e$, the formula~(\ref{eq.2.71}) reduces to~(\ref{eq.2.66}).

In addition to the leptonic contribution, we also need to consider the hadronic contribution to the vacuum polarization. The total amplitude, ${\mathcal M}_{\text{vac}}$, which includes both the leptonic and hadronic parts, is expressed through the so-called photon polarization operator, ${\mathcal P} (q^2)$, as
\begin{equation}
{\mathcal M}_{\text{vac}} = {\mathcal P} (q^2) \, {\mathcal M}_{\text{Born}}, \label{eq.2.73}
\end{equation}
which implies
\begin{equation}
\delta_{\text{vac}} = 2\re{{\mathcal P} (q^2)}. \label{eq.2.74}
\end{equation}

\begin{figure}[t]
\centering
\includegraphics[width=0.8\textwidth]{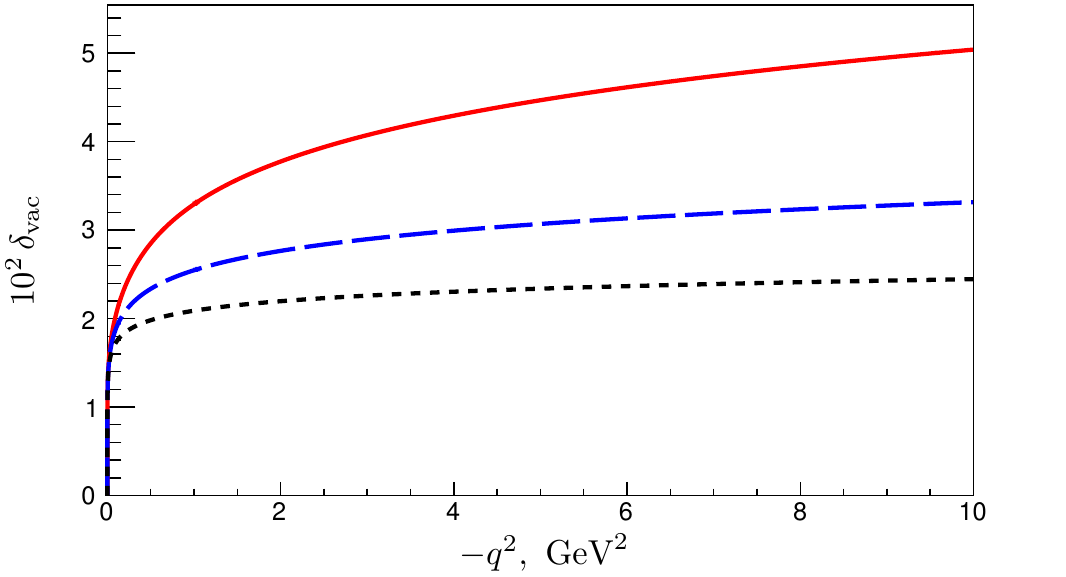}
\caption{Different contributions to the vacuum polarization correction~$\delta_{\text{vac}}$ as functions of~$q^2$: the contribution~$\delta_{\text{vac}}^{e}$~(\ref{eq.2.66}) from electron--positron loops only (black short-dashed line), the full leptonic contribution~$\delta_{\text{vac}}^{e} + \delta_{\text{vac}}^{\mu} + \delta_{\text{vac}}^{\tau}$ (\ref{eq.2.71}) (blue long-dashed line), and the full vacuum polarization correction $\delta_{\text{vac}}$ (\ref{eq.2.74}) (red solid line).}
\label{Fig4}
\end{figure}

The hadronic part of~${\mathcal P} (q^2)$ is difficult to calculate theoretically, but it can be reliably extracted from the experimental data for the cross section of the annihilation process $e^+ e^- \rightarrow$~\emph{hadrons}. In our event generator, we use the results of a global analysis carried out by F.\,V.~Ignatov~\cite{vpl, EPJC.66.585}. Figure~\ref{Fig4} shows the different contributions to the vacuum polarization correction as functions of~$q^2$.

We do not assume that $-q^2 \gg m^2$ when accounting for the vacuum polarization using the formula~(\ref{eq.2.74}). However, this approximation has been used in the derivation of the lepton vertex correction and thus the expressions~(\ref{eq.2.55}) and~(\ref{eq.2.67}) require refinement in the case when $-q^2 \approx m^2$. Note that in this case the amplitude~${\mathcal M}_{\text{vert}}^{\ell}$ is not expressed in terms of the Born amplitude~${\mathcal M}_{\text{Born}}$, unlike in formula~(\ref{eq.2.55}). This issue is discussed, for example, in~\cite{arXiv:1207.6651}.

Finally, we can refine the expressions for the amplitudes~${\mathcal M}_{\text{box}}^{\text{MTs}}$ and~${\mathcal M}_{\text{xbox}}^{\text{MTs}}$ describing the soft parts of the TPE contribution. Maximon and Tjon~\cite{PRC.62.054320} obtained the following formulas using a less drastic approximation than that employed by Mo and Tsai:
\begin{gather}
{\mathcal M}_{\text{box}}^{\text{MTj}} = z \frac{\alpha}{\pi} \, \frac{E_{\ell}}{|{\boldsymbol \ell}|} \, \ln{\!\left(\frac{E_{\ell} + |{\boldsymbol \ell}|}{m}\right)} \ln{\!\left(\frac{-q^2}{\lambda^2}\right)} \, {\mathcal M}_{\text{Born}}, \label{eq.2.75} \\
{\mathcal M}_{\text{xbox}}^{\text{MTj}} = -z \frac{\alpha}{\pi} \, \frac{E_{\ell}'}{|{\boldsymbol \ell}'|} \, \ln{\!\left(\frac{E_{\ell}' + |{\boldsymbol \ell}'|}{m}\right)} \ln{\!\left(\frac{-q^2}{\lambda^2}\right)} \, {\mathcal M}_{\text{Born}}. \label{eq.2.76}
\end{gather}
The amplitudes~(\ref{eq.2.75}) and~(\ref{eq.2.76}) by Maximon and Tjon, as well as the amplitudes~(\ref{eq.2.60}) and~(\ref{eq.2.58}) by Mo and Tsai, are infrared-divergent, but the difference $\bigl({\mathcal M}_{\text{box}}^{\text{MTj}} + {\mathcal M}_{\text{xbox}}^{\text{MTj}} - {\mathcal M}_{\text{box}}^{\text{MTs}} - {\mathcal M}_{\text{xbox}}^{\text{MTs}}\bigr)$ is finite and gives the following addition to~$\delta_{\text{virt}}$~\cite{PPNP.66.782}:
\begin{equation}
\delta_{2\gamma}' = z \frac{\alpha}{\pi} \left[\ln{\frac{E_{\ell}}{E_{\ell}'}} \, \ln{\frac{q^4}{4 M^2 E_{\ell} E_{\ell}'}} + 2\dilog{\!\left(1 - \frac{M}{2 E_{\ell}}\right)} - 2\dilog{\!\left(1 - \frac{M}{2 E_{\ell}'}\right)}\right], \label{eq.2.77}
\end{equation}
where the approximation $E_{\ell}, E_{\ell}' \gg m$ is assumed and the function~$\dilog$ is the dilogarithm (Spence function) defined as
\begin{equation}
\dilog{(x)} = -\int\limits_0^x \frac{\ln{|1 - y|}}{y} \, {\rm d} y. \label{eq.2.78}
\end{equation}

\begin{figure}[t]
\centering
\includegraphics[width=0.8\textwidth]{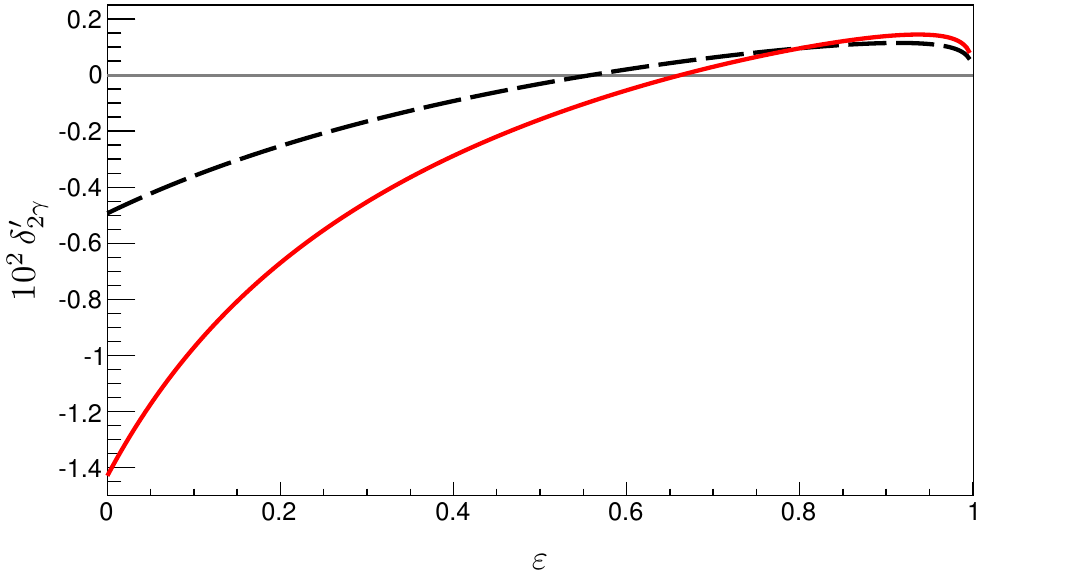}
\caption{Dependence of~$\delta_{2\gamma}'$ on~$\varepsilon$ for two fixed values of the four-momentum transfer squared: $-q^2 = 1~\text{GeV}^2$ (black dashed line) and $-q^2 = 5~\text{GeV}^2$ (red solid line). Both lines are shown for the case of electron--proton scattering ($z = -1$).}
\label{Fig5}
\end{figure}

Figure~\ref{Fig5} shows the dependence of~$\delta_{2\gamma}'$ on~$\varepsilon$ for two fixed values of the four-momentum transfer squared ($1$ and $5~\text{GeV}^2$). It can be seen that~$\delta_{2\gamma}'$ has a significant dependence on $\varepsilon$, so this simple TPE correction already affects the Rosenbluth measurements of the proton electromagnetic form factors (see~\cite{PPNP.66.782} for further discussion).

To account for all the additional corrections mentioned, the formula~(\ref{eq.2.64}) should be modified as
\begin{equation}
\delta_{\text{virt}} = \delta_{\text{vac}} + \delta_{\text{vert}} + \delta_{2\gamma}', \label{eq.2.79}
\end{equation}
where~$\delta_{\text{vac}}$ is now given by the expression~(\ref{eq.2.74}) and $\delta_{2\gamma}'$ is given by~(\ref{eq.2.77}).

\section{Radiative corrections in TPE measurements}
\label{S_3}
\setcounter{equation}{0}

In this section, we discuss the practical aspects of taking into account the standard radiative corrections for the experiments~\cite{nucl-ex.0408020, NPBPS.225-227.216, AIPConfProc.1160.19, NIMA.741.1, AIPConfProc.1160.24, PRC.88.025210} aimed at measuring the hard TPE effects in elastic lepton--proton scattering. These experiments make a precise comparison between the elastic electron--proton and positron--proton scattering cross sections, $\sigma (e^- p)$ and $\sigma (e^+ p)$. There are two natural dimensionless combinations of $\sigma (e^- p)$ and~$\sigma (e^+ p)$ to quantitatively characterize the difference between these cross sections~--- the cross section ratio and asymmetry, $R$ and $A$, defined as
\begin{equation}
R = \frac{\sigma (e^+ p)}{\sigma (e^- p)}, \qquad A = \frac{\sigma (e^+ p) - \sigma (e^- p)}{\sigma (e^+ p) + \sigma (e^- p)}, \label{eq.3.1}
\end{equation}
and therefore simply related to each other by
\begin{equation}
R = \frac{1 + A}{1 - A}, \qquad A = \frac{R - 1}{R + 1}. \label{eq.3.2}
\end{equation}
For historical reasons, the ratio~$R$ is commonly used, though the asymmetry~$A$ is more convenient for analysis because it contains only charge-odd terms in the numerator and only charge-even terms in the denominator. We will consider both of these quantities.

In the discussed TPE experiments, the numbers of events of elastic $e^- p$ and $e^+ p$ scattering, $N_{\text{meas}}^{-}$ and $N_{\text{meas}}^{+}$, are measured under very similar experimental conditions. In this case, factors such as the acceptances of the detectors, detection efficiencies, target thicknesses and beam current integrals can be kept almost the same for the cases of electron--proton and positron--proton scattering, so they do not affect the measured cross section ratio or asymmetry. Therefore, it is much easier to measure the quantities (\ref{eq.3.1}) precisely than to measure the absolute cross sections $\sigma (e^- p)$ and $\sigma (e^+ p)$ with the same relative accuracy.

The ultimate goal of these experiments is to measure the hard TPE contribution,
\begin{equation}
\delta_{2\gamma} = \frac{2\re{\bigl({\mathcal M}_{\text{Born}}^{\dagger} {\mathcal M}_{2\gamma}^{\text{hard}}\bigr)}}{|{\mathcal M}_{\text{Born}}|^2}, \label{eq.3.3}
\end{equation}
to the cross section~(\ref{eq.2.2}) of elastic lepton--proton scattering in some kinematic region. It should be recalled that the separation of the amplitude ${\mathcal M}_{2\gamma}$ into soft and hard parts, ${\mathcal M}_{2\gamma}^{\text{soft}}$ and ${\mathcal M}_{2\gamma}^{\text{hard}}$, is not unique and is thus important to make explicit. For example, the contribution~$\delta_{2\gamma}'$, given by~(\ref{eq.2.77}), is soft and included in the standard radiative corrections according to Maximon and Tjon, while in the approach of Mo and Tsai it should be considered as hard and therefore contained in~(\ref{eq.3.3}).

The quantity~$\delta_{2\gamma}$ is charge-odd and depends on the kinematics of elastic scattering. It can be represented as a function of two independent kinematic variables, for example, $q^2$ and $\varepsilon$. Precise measurements of~$\delta_{2\gamma}$ at several kinematic points will test the available theoretical calculations and help to understand whether the hard TPE effects explain the observed discrepancy between the two methods for measuring the ratio~$G_E / G_M$.

It is important to understand how to express the quantity (\ref{eq.3.3}) through the measured numbers~$N_{\text{meas}}^-$ and~$N_{\text{meas}}^+$. In other words, we want to know how to properly take into account the processes of the lowest order in~$\alpha$ giving the standard radiative corrections in the considered measurements. The procedure for taking into account these corrections consists of generating events of elastic $e^- p$ and~$e^+ p$ scattering and conducting a Monte Carlo simulation of the detector response to these events using, for example, the Geant4 toolkit. This simulation should accurately reproduce the experimental conditions and the real procedure of event selection. As the simulation result, we obtain the numbers of $e^- p$ and $e^+ p$ scattering events, $N_{\text{sim}}^-$ and $N_{\text{sim}}^+$, which can also be combined as the asymmetry~$A_{\text{sim}}$.

If we consider only the first-order radiative corrections in accordance with~(\ref{eq.2.1}) and take into account the parity of each amplitude with respect to~$z$, then we can write the following expressions for the experimentally measured and simulated asymmetries, $A_{\text{meas}}$ and $A_{\text{sim}}$:
\begin{align}
A_{\text{meas}} &= \frac{N_{\text{meas}}^{+} - N_{\text{meas}}^{-}}{N_{\text{meas}}^{+} + N_{\text{meas}}^{-}} \nonumber \\
{} &= z \, \frac{2\re{\bigl[{\mathcal M}_{\text{Born}}^{\dagger} \left({\mathcal M}_{2\gamma}^{\text{soft}} + {\mathcal M}_{2\gamma}^{\text{hard}}\right)\bigr]} + 2\re{\bigl({\mathcal M}_{\text{brems}}^{\ell \dagger} {\mathcal M}_{\text{brems}}^p\bigr)}}{|{\mathcal M}_{\text{Born}}|^2 + 2\re{\bigl({\mathcal M}_{\text{Born}}^{\dagger} {\mathcal M}_{\text{virt}}\bigr)} + \bigl|{\mathcal M}_{\text{brems}}^{\ell}\bigr|^2 + \bigl|{\mathcal M}_{\text{brems}}^p\bigr|^2}, \label{eq.3.4} \\
A_{\text{sim}} &= \frac{N_{\text{sim}}^{+} - N_{\text{sim}}^{-}}{N_{\text{sim}}^{+} + N_{\text{sim}}^{-}} \nonumber \\
{} &= z \, \frac{2\re{\bigl({\mathcal M}_{\text{Born}}^{\dagger} {\mathcal M}_{2\gamma}^{\text{soft}}\bigr)} + 2\re{\bigl({\mathcal M}_{\text{brems}}^{\ell \dagger} {\mathcal M}_{\text{brems}}^p\bigr)}}{|{\mathcal M}_{\text{Born}}|^2 + 2\re{\bigl({\mathcal M}_{\text{Born}}^{\dagger} {\mathcal M}_{\text{virt}}\bigr)} + \bigl|{\mathcal M}_{\text{brems}}^{\ell}\bigr|^2 + \bigl|{\mathcal M}_{\text{brems}}^p\bigr|^2}, \label{eq.3.5}
\end{align}
where ${\mathcal M}_{\text{virt}} = {\mathcal M}_{\text{vac}} + {\mathcal M}_{\text{vert}}^{\ell} + {\mathcal M}_{\text{vert}}^p$. Taking the difference $A_{\text{meas}} - A_{\text{sim}}$ we can reduce the interference terms due to soft TPE and bremsstrahlung in the numerator, while the denominator remains unchanged. After that, it is easy to see that the desired quantity~(\ref{eq.3.3}) can be expressed as
\begin{equation}
\delta_{2\gamma} = z \left(\frac{N_{\text{meas}}^{+} - N_{\text{meas}}^{-}}{N_{\text{meas}}^{+} + N_{\text{meas}}^{-}} - \frac{N_{\text{sim}}^{+} - N_{\text{sim}}^{-}}{N_{\text{sim}}^{+} + N_{\text{sim}}^{-}}\right) \frac{N_{\text{sim}}^{+} + N_{\text{sim}}^{-}}{2N_{\text{sim}}^{0}}, \label{eq.3.6}
\end{equation}
where $N_{\text{sim}}^0$ is the number of events corresponding to the value of~$|{\mathcal M}_{\text{Born}}|^2$, i.e., in the absence of any radiative corrections. This number can be obtained in a Monte Carlo simulation using the elastic events generated in accordance with the Rosenbluth formula~(\ref{eq.2.4}). It is assumed that the numbers of simulated events of all three types correspond to the same value of integrated luminosity. Let us also note that the factor $(N_{\text{sim}}^{+} + N_{\text{sim}}^{-}) / 2 N_{\text{sim}}^0$ can be neglected in many practical cases.

One may prefer to present the experimental results in the form of TPE ratio or asymmetry, $R_{2\gamma}$ or $A_{2\gamma}$. The expressions for them directly follow from (\ref{eq.2.1}), (\ref{eq.3.1}), and~(\ref{eq.3.3}) if we assume that only the amplitudes ${\mathcal M}_{\text{Born}}$ and ${\mathcal M}_{2\gamma}^{\text{hard}}$ contribute to the cross sections $\sigma (e^- p)$ and $\sigma (e^+ p)$
\begin{equation}
R_{2\gamma} = \frac{1 + z \delta_{2\gamma}}{1 - z \delta_{2\gamma}}, \qquad A_{2\gamma} = z \delta_{2\gamma}. \label{eq.3.7}
\end{equation}

\section{Description of the event generator}
\label{S_4}
\setcounter{equation}{0}

\texttt{ESEPP} is written in the C++ programming language using some ROOT~\cite{ROOT} classes. Its source code is publicly available on the GitHub repository~\cite{ESEPP} and can be used as a basis for future developments. The source distribution of \texttt{ESEPP} includes detailed information on compiling the code, the input parameters, and the format of output files. In this section, let us briefly describe the algorithm implemented in the generator.

\texttt{ESEPP} generates events of two types~--- `elastic' ($\ell^{\pm} p \rightarrow \ell^{\pm} p$) and `inelastic' ($\ell^{\pm} p \rightarrow \ell^{\pm} p \, \gamma$). It is important to understand that first-order bremsstrahlung is taken into account in both cases, and the separation between these two types of events is based on the energy of the emitted photon. If this energy does not exceed the cut-off value $E_{\gamma}^{\text{cut}}$ (which can be set to, for example, $1~\text{MeV}$ for the experiments of interest), then the scattering process can be effectively considered as elastic since the experiment can not distinguish it from the purely elastic process. In this case, it is not necessary to consider such soft photons in the Monte Carlo simulation, so we can analytically integrate the cross section of the first-order bremsstrahlung process $\ell^{\pm} p \rightarrow \ell^{\pm} p \, \gamma$ over all directions of the emitted photon and over its energy in the range~$E_{\gamma} < E_{\gamma}^{\text{cut}}$. Moreover, we can confidently use the primary soft-photon approximation to perform this integration, because it works well for very low-energy photons. The procedure of integration is described in section~\ref{Ss_2.3} and appendix~\hyperlink{app1}{A}. To generate elastic events, we use the formula~(\ref{eq.2.63}), where all virtual-photon radiative corrections are included and the infrared divergences are canceled out analytically.

Inelastic events can be generated using any one of three different models within the soft-photon approximation (`primary', `modified', and `improved'~--- discussed in section~\ref{Ss_2.3}) or using the accurate QED calculation described in section~\ref{Ss_2.4}. In the latter case, formula (\ref{eq.2.39}) is used, a detailed derivation of which is given in appendix~\hyperlink{app2}{B}. The differential cross section (\ref{eq.2.39}) is expressed through the squared amplitude $|{\mathcal M}_{\text{brems}}|^2$ of the process $\ell^{\pm} p \rightarrow \ell^{\pm} p \, \gamma$, which is given by the formulas~(\ref{eq.2.40}) and (\ref{eq.2.45})--(\ref{eq.2.53}). To calculate the tensor contractions in (\ref{eq.2.45}), (\ref{eq.2.48}), and (\ref{eq.2.51}) we used the \texttt{FeynCalc} package~\cite{FeynCalc} for Mathematica. Details of these calculations can be found in~\cite{ESEPP}. The resulting formulas are expressed through the scalar products $p_i \cdot p_j$ of the four-momenta $\ell$, $\ell'$, $p$, $p'$, and~$k$. We used the \texttt{cform} command to convert these cumbersome expressions obtained in Mathematica to C++ code. Let us note that we did not neglect the lepton mass in the derivation of the cross section (\ref{eq.2.39}) and in the calculation of~$|{\mathcal M}_{\text{brems}}|^2$.

We choose as the basic kinematic variables the following: the lepton scattering angle, the energy of the bremsstrahlung photon, and its polar and azimuthal angles. The fixed energy of the incident lepton is also assumed to be known. Then, the remaining kinematic parameters of the final-state particles can be calculated (up to an arbitrary angle of rotation around the axis defined by the incident lepton) as described in section~\ref{Ss_2.2}. The expressions given there were also obtained without neglecting the lepton mass.

We have to use definite values of the form factors~$G_E (q^2)$ and~$G_M (q^2)$ to calculate the differential cross section ${\rm d} \sigma_{\text{Born}} / {\rm d} \Omega_{\ell}$ and the squared amplitude $|{\mathcal M}_{\text{brems}}|^2$. In the generator, we have implemented several different models for these form factors, described in section~\ref{Ss_2.1}. We also made it possible to use the parametrization (\ref{eq.2.12}) and set arbitrary values for the coefficients $a_{i1}$, $b_{i1}$, $b_{i2}$, and $b_{i3}$. When using the generator, one should be aware of the possible model dependence of simulation results on the parametrization used for the proton form factors.

An unweighted event generator, such as \texttt{ESEPP}, randomly generates a certain number of $n$-dimensional vectors (where $n$ is the number of basic kinematic variables needed to completely describe the event kinematics) according to a given probability distribution~--- the differential cross section. This procedure is based on the use of a generator of pseudorandom numbers uniformly distributed between $0$ and~$1$. For elastic events, there is only one basic kinematic variable~--- the lepton scattering angle, so the probability distribution, described by the formula~(\ref{eq.2.63}), is one-dimensional. In this case, one can simply use rejection sampling (the acceptance-rejection method) to generate events. However, in the case of inelastic events, there are $n = 4$ basic kinematic variables ($\theta_{\ell}$, $E_{\gamma}$, $\theta_{\gamma}$, and $\varphi_{\gamma}$), which form the so-called four-dimensional phase space. The probability distribution is also four-dimensional and is described by the formulas~(\ref{eq.2.28}) or~(\ref{eq.2.39}). It is obvious that this distribution has peaks associated with each kinematic variable (for example, the cross section is usually higher for the lowest scattering angles, for the lowest photon energies, and when the photon is emitted along either the incident or scattered lepton). Now, simple rejection sampling becomes very inefficient to generate events. 

For this reason, we use \texttt{mFOAM}~\cite{CPC.152.55, CPC.177.441}, the general-purpose self-adapting Monte Carlo event generator and integrator, which is built into the ROOT package~\cite{ROOT} as the \texttt{TFoam} class. \texttt{mFOAM} divides the phase space into many hyper-rectangular cells (`foam'), which are more dense around the peaks in the distribution. It allows us to greatly improve the rejection efficiency for each single cell. \texttt{mFOAM} is also able to numerically integrate a given (unnormalized) probability distribution over the entire phase space. We use this feature to calculate the integrated cross sections of the considered processes. It is necessary in order to determine how many events of each of the four types ($\ell^- p \rightarrow \ell^- p$, $\ell^- p \rightarrow \ell^- p \, \gamma$, $\ell^+ p \rightarrow \ell^+ p$, and $\ell^+ p \rightarrow \ell^+ p \, \gamma$) we need to generate. The obvious requirement is that the integrated luminosity (which is the ratio of the number of events to the integrated cross section) must be the same for each of these processes, while the total number of events is one of the input parameters specified by a user.

\section{Summary}
\label{S_5}
\setcounter{equation}{0}

A renewed interest has been recently shown in experiments on unpolarized elastic scattering of charged leptons on protons. Any such experiment nowadays requires performing a careful Monte Carlo simulation, which helps to optimize the detector configuration, better understand the sources of systematic uncertainties of the measurement, and accurately take into account the radiative corrections to the measured cross sections. To facilitate these simulations, the \texttt{ESEPP} generator produces unweighted events with the kinematic parameters of all final-state particles, taking into account the lowest-order QED radiative corrections to the Rosenbluth cross section without using the common soft-photon or ultrarelativistic approximations. The generator can be useful for several significant ongoing and planned experiments, such as studying TPE effects, measuring the electromagnetic form factors and charge radius of the proton, and searching for new physics in elastic lepton--proton scattering. The source code for \texttt{ESEPP} is publicly available and can be used as a basis for future developments.

The underlying theoretical formulas and calculations for the generator have been described in this paper. The practical aspects of taking into account the standard radiative corrections in TPE measurements have been specifically discussed. We have also provided a description of the basic algorithm implemented in the generator.

Our approach to considering first-order bremsstrahlung is the most direct. It combines analytical and numerical calculations of the cross section of the process ${\ell}^{\pm} p \rightarrow {\ell}^{\pm} p \, \gamma$. We introduce the cut-off energy of bremsstrahlung photons, $E_{\gamma}^{\text{cut}}$, which separates the regions of analytical and numerical integration of the cross section over the phase space. Analytical integration is necessary in order to properly cancel out the infrared divergences, but it can only be performed using the soft-photon approximation. We use this approximation only for the photons with energies below $E_{\gamma}^{\text{cut}}$, when its validity is unquestionable. At the same time, numerical calculation and integration of the cross section allow us to not use the soft-photon or ultrarelativistic approximations at all. To do this, we use the computer algebra toolkits, Mathematica and \texttt{FeynCalc}, to calculate the squared amplitude of the process, and the self-adapting Monte Carlo generator \texttt{mFOAM} to integrate the cross section numerically and to generate unweighted events.

\label{ack.1}
\pdfbookmark[1]{Acknowledgments}{ack.1}
\section*{Acknowledgments}

The authors would like to thank V.\,F.~Dmitriev, A.~Gasparian, and A.\,V.~Grabovsky for their interest in this work and helpful discussions. The attention that was paid to our work by L.\,M.~Barkov (1928~--~2013) was also motivating for us. The corresponding author is very grateful to R.\,L.~Russell for advice on improving the generator. This research was supported in part by the Ministry of Education and Science of the Russian Federation (Project 14.B37.21.1181), the Russian Foundation for Basic Research (Grants 12-02-33140 and 13-02-01023), and the US National Science Foundation (Award PHY-0855543). Two of the authors (A.L.F.\ and R.E.G.) acknowledge the Dynasty Foundation for financial support.

\label{appa.1}
\pdfbookmark[1]{Appendix A}{appa.1}
\hypertarget{app1}{\section*{Appendix A. Calculation of $B \bigl(p_i, p_j, E_{\gamma}^{\text{cut}}\bigr)$}}
\renewcommand{\theequation}{A.\arabic{equation}}
\setcounter{equation}{0}

To calculate $B \bigl(p_i, p_j, E_{\gamma}^{\text{cut}}\bigr)$, we need to perform the following integration over all directions of the bremsstrahlung photon and over its energy in the range~$E_{\gamma} < E_{\gamma}^{\text{cut}}$
\begin{equation}
B \bigl(p_i, p_j, E_{\gamma}^{\text{cut}}\bigr) = \frac{1}{8\pi^2} \int\limits_{E_{\gamma} < E_{\gamma}^{\text{cut}}} \frac{{\rm d}^3 k}{E_{\gamma}} \, \frac{p_i \cdot p_j}{(k \cdot p_i) (k \cdot p_j)}. \label{eq.A.1}
\end{equation}
Let us show how to do this within the primary soft-photon approximation (see~\cite{PRC.64.054610, Itzykson&Zuber} also). This approximation assumes that only the four-momentum~$k$ varies in the integration, while the other four-momenta are constant. We also assume that the photon has a nonzero mass~$\lambda$ and thus its full energy is $E_{\gamma} = \sqrt{|{\mathbf k}|^2 + \lambda^2}$. Then, taking into account that~${\rm d}^3 k = |{\mathbf k}|^2 {\rm d} |{\mathbf k}| \, {\rm d} \Omega_{\gamma}$, we obtain
\begin{equation}
B \bigl(p_i, p_j, E_{\gamma}^{\text{cut}}\bigr) = \frac{p_i \cdot p_j}{8\pi^2} \int\limits_0^{\sqrt{(E_{\gamma}^{\text{cut}})^2 - \lambda^2}} \frac{|{\mathbf k}|^2 {\rm d} |{\mathbf k}|}{\sqrt{|{\mathbf k}|^2 + \lambda^2}} \int \frac{{\rm d} \Omega_{\gamma}}{(k \cdot p_i) (k \cdot p_j)}. \label{eq.A.2}
\end{equation}

Using the Feynman parametrization, we can make the transformation
\begin{equation}
\frac{1}{(k \cdot p_i) (k \cdot p_j)} = \int\limits_0^1 \frac{{\rm d} x}{\bigl[(k \cdot p_i) x + (k \cdot p_j) (1 - x)\bigr]^2} = \int\limits_0^1 \frac{{\rm d} x}{{(k \cdot p_x)}^2}, \label{eq.A.3}
\end{equation}
where the four-momentum~$p_x$ is defined by~(\ref{eq.2.32}). Therefore,
\begin{gather}
\int \frac{{\rm d} \Omega_{\gamma}}{(k \cdot p_i) (k \cdot p_j)} = \int {\rm d} \Omega_{\gamma} \int\limits_0^1 \frac{{\rm d} x}{{(k \cdot p_x)}^2} = 2\pi \int\limits_{0}^{\pi} \sin{\theta} \, {\rm d} \theta \int\limits_0^1 \frac{{\rm d} x}{\bigl(E_{\gamma} p_x^0 - |{\mathbf k}| \, |{\mathbf p}_x| \cos{\theta}\bigr)^2} \nonumber \\
{} = -2 \pi \int\limits_0^1 {\rm d} x \int\limits_{1}^{-1} \frac{{\rm d} (\cos{\theta})}{\bigl(E_{\gamma} p_x^0 - |{\mathbf k}| \, |{\mathbf p}_x| \cos{\theta}\bigr)^2} = 4\pi \int\limits_0^1 \frac{{\rm d} x}{\bigl(E_{\gamma} p_x^0\bigr)^2 - |{\mathbf k}|^2 |{\mathbf p}_x|^2}, \label{eq.A.4}
\end{gather}
where $\theta$ is the angle between the vectors ${\mathbf k}$ and~${\mathbf p}_x$. Substituting this expression into~(\ref{eq.A.2}) and using the identity $E_{\gamma} = \sqrt{|{\mathbf k}|^2 + \lambda^2}$, we obtain
\begin{equation}
B \bigl(p_i, p_j, E_{\gamma}^{\text{cut}}\bigr) = \frac{p_i \cdot p_j}{2\pi} \int\limits_0^1 {\rm d} x \! \! \int\limits_0^{\sqrt{(E_{\gamma}^{\text{cut}})^2 - \lambda^2}} \! \frac{1}{\sqrt{|{\mathbf k}|^2 + \lambda^2}} \, \frac{|{\mathbf k}|^2 {\rm d} |{\mathbf k}|}{|{\mathbf k}|^2 \left[\bigl(p_x^0\bigr)^2 - |{\mathbf p}_x|^2\right] + \bigl(\lambda \, p_x^0\bigr)^2}. \label{eq.A.5}
\end{equation}
Further, the trigonometric substitution~$|{\mathbf k}| = \lambda \tan{\psi}$ gives
\begin{equation}
B \bigl(p_i, p_j, E_{\gamma}^{\text{cut}}\bigr) = \frac{p_i \cdot p_j}{2\pi} \int\limits_0^1 {\rm d} x \int\limits_0^{\arctan{\sqrt{(E_{\gamma}^{\text{cut}} / \lambda)^2 - 1}}} \frac{\sin{\psi} \, \tan{\psi} \, {\rm d} \psi}{\bigl(p_x^0\bigr)^2 - |{\mathbf p}_x|^2 \sin^2{\psi}}. \label{eq.A.6}
\end{equation}

Let us denote the inner integral in~(\ref{eq.A.6}) by the symbol~$I$ and consider it separately. Performing another trigonometric substitution, $\sin{\psi} = t$, we obtain
\begin{equation}
I = \int\limits_0^{\sqrt{1 - {(\lambda / E_{\gamma}^{\text{cut}})}^2}} \frac{t^2 \, {\rm d} t}{\bigl(1 - t^2\bigr) \left[\bigl(p_x^0\bigr)^2 - |{\mathbf p}_x|^2 \, t^2\right]}. \label{eq.A.7}
\end{equation}
Then, applying the algebraic decomposition
\begin{gather}
\frac{t^2}{\bigl(1 - t^2\bigr) \Bigl[{\bigl(p_x^0\bigr)}^2 - |{\mathbf p}_x|^2 \, t^2\Bigr]} \nonumber \\
{} = \frac{1}{2\Bigl[{\bigl(p_x^0\bigr)}^2 - |{\mathbf p}_x|^2\Bigr]} \left(\frac{p_x^0}{|{\mathbf p}_x| \, t - p_x^0} - \frac{p_x^0}{|{\mathbf p}_x| \, t + p_x^0} - \frac{1}{t - 1} + \frac{1}{t + 1}\right), \label{eq.A.8}
\end{gather}
we can find that
\begin{gather}
I = \frac{1}{2\Bigl[{\bigl(p_x^0\bigr)}^2 - |{\mathbf p}_x|^2\Bigr]} \left(\frac{p_x^0}{|{\mathbf p}_x|} \ln{\bigl||{\mathbf p}_x| \, t - p_x^0\bigr|} \right. \nonumber \\
{} \left. - \frac{p_x^0}{|{\mathbf p}_x|} \ln{\bigl||{\mathbf p}_x| \, t + p_x^0\bigr|} - \ln{|t - 1|} + \ln{|t + 1|}\right)\left. \rule{0mm}{6mm} \right|_0^{\sqrt{1 - {(\lambda / E_{\gamma}^{\text{cut}})}^2}}. \label{eq.A.9}
\end{gather}
Now, using the approximation~$\lambda \ll E_{\gamma}^{\text{cut}}$ and the fact that~${\bigl(p_x^0\bigr)}^2 - |{\mathbf p}_x|^2 = p_x^2$, we obtain
\begin{equation}
I = \frac{1}{2p_x^2} \left(\frac{p_x^0}{|{\mathbf p}_x|} \ln{\frac{p_x^0 - |{\mathbf p}_x|}{p_x^0 + |{\mathbf p}_x|}} + \ln{\frac{4 {\bigl(E_{\gamma}^{\text{cut}}\bigr)}^2}{\lambda^2}}\right), \label{eq.A.10}
\end{equation}
which allows us to write the desired expression for~$B \bigl(p_i, p_j, E_{\gamma}^{\text{cut}}\bigr)$:
\begin{equation}
B \bigl(p_i, p_j, E_{\gamma}^{\text{cut}}\bigr) = \frac{p_i \cdot p_j}{4\pi} \int\limits_0^1 \frac{{\rm d} x}{p_x^2} \left(\ln{\frac{4 \bigl(E_{\gamma}^{\text{cut}}\bigr)^2}{p_x^2}} + \frac{p_x^0}{|{\mathbf p}_x|} \ln{\frac{p_x^0 - |{\mathbf p}_x|}{p_x^0 + |{\mathbf p}_x|}} + \ln{\frac{p_x^2}{\lambda^2}}\right). \label{eq.A.11}
\end{equation}

\label{appb.1}
\pdfbookmark[1]{Appendix B}{appb.1}
\hypertarget{app2}{\section*{Appendix B. Derivation of a general expression for the differential cross section of the process ${\ell}^{\pm} p \rightarrow {\ell}^{\pm} p \, \gamma$}}
\renewcommand{\theequation}{B.\arabic{equation}}
\setcounter{equation}{0}

The fully-differential cross section for the process $\ell^{\pm} p \rightarrow \ell^{\pm} p \, \gamma$ can be written as~\cite{Berestetskii}
\begin{equation}
{\rm d} \sigma_{\text{brems}} = (2\pi)^4 \, \delta^{(4)} (P_i - P_f) \, \frac{1}{4 I} \, |{\mathcal M}_{\text{brems}}|^2 \, \frac{{\rm d}^3 \ell'}{(2\pi)^3 \, 2 E_{\ell}'} \, \frac{{\rm d}^3 p'}{(2\pi)^3 \, 2 E_p} \, \frac{{\rm d}^3 k}{(2\pi)^3 \, 2 E_{\gamma}}, \label{eq.B.1}
\end{equation}
where
\begin{gather}
P_i = \ell + p, \qquad P_f = \ell' + p' + k, \label{eq.B.2} \\
\delta^{(4)} (P_i - P_f) = \delta (E_{\ell} + M - E_{\ell}' - E_p - E_{\gamma}) \, \delta^{(3)} ({\boldsymbol \ell} - {\boldsymbol \ell}' - {\mathbf p}' - {\mathbf k}), \label{eq.B.3} \\
I = \sqrt{(\ell \cdot p)^2 - m^2 M^2} = M |{\boldsymbol \ell}|. \label{eq.B.4}
\end{gather}

Integrating~(\ref{eq.B.1}) with respect to~$p'$ and using the identity ${\rm d}^3 p_i = |{\mathbf p}_i| \, E_i \, {\rm d} E_i \, {\rm d} \Omega_i$, we obtain
\begin{equation}
{\rm d} \sigma_{\text{brems}} = \frac{1}{(4\pi)^5} \, \delta (E_{\ell} + M - E_{\ell}' - E_p - E_{\gamma}) \, \frac{|{\mathcal M}_{\text{brems}}|^2}{M |{\boldsymbol \ell}|} \, \frac{|{\boldsymbol \ell}'| |{\mathbf k}|}{E_p} \, {\rm d} E_{\ell}' \, {\rm d} E_{\gamma} \, {\rm d} \Omega_{\ell} \, {\rm d} \Omega_{\gamma}. \label{eq.B.5}
\end{equation}
For the integration with respect to~$E_{\ell}'$, we will use the well-known representation of the delta function of the argument being itself a function of the independent variable~$x$:
\begin{equation}
\delta \left[f (x)\right] = \sum_{i=1}^k \frac{\delta (x - x_i)}{\bigl|{\rm d} f(x_i) / {\rm d} x \bigr|}, \label{eq.B.6}
\end{equation}
where $f (x_i) = 0$ and $i = 1, 2, \ldots, k$. In our case,
\begin{equation}
f (E_{\ell}') = E_{\ell} + M - E_{\ell}' - E_{\gamma} - \sqrt{({\boldsymbol \ell} - {\mathbf k})^2 - 2({\boldsymbol \ell} - {\mathbf k}) \cdot {\boldsymbol \ell}' + {E_{\ell}'}^2 - m^2 + M^2}, \label{eq.B.7}
\end{equation}
and the equation $f(E_{\ell}') = 0$ has at most two roots listed in~(\ref{eq.2.24}). Differentiation of~(\ref{eq.B.7}) with respect to~$E_{\ell}'$ gives
\begin{equation}
\frac{{\rm d} f(E_{\ell}')}{{\rm d} E_{\ell}'} = -1 - \frac{E_{\ell}'}{E_p} \left[1 - \frac{({\boldsymbol \ell} - {\mathbf k}) \cdot {\boldsymbol \ell}'}{|{\boldsymbol \ell}'|^2}\right] = \frac{A E_{\ell}' - B |{\boldsymbol \ell}'|}{E_p |{\boldsymbol \ell}'|}, \label{eq.B.8}
\end{equation}
where the coefficients $A$ and $B$ are defined by~(\ref{eq.2.20})--(\ref{eq.2.22}).

Finally, substituting (\ref{eq.B.6}) and~(\ref{eq.B.8}) into~(\ref{eq.B.5}) and integrating the result with respect to~$E_{\ell}'$, we obtain
\begin{equation}
\frac{{\rm d} \sigma_{\text{brems}}}{{\rm d} E_{\gamma} \, {\rm d} \Omega_{\gamma} \, {\rm d} \Omega_{\ell}} = \frac{1}{(4\pi)^5} \, \frac{1}{M |{\boldsymbol \ell}|} \sum_{E_{\ell}'} \frac{E_{\gamma} \, |{\boldsymbol \ell}'|^2 \, |{\mathcal M}_{\text{brems}}|^2}{\bigl|A E_{\ell}' - B |{\boldsymbol \ell}'|\bigr|}, \label{eq.B.9}
\end{equation}
where $E_{\ell}'$ and~$|{\boldsymbol \ell}'|$ can be expressed in terms of $E_{\ell}$, $\theta_{\ell}$, $E_{\gamma}$, $\theta_{\gamma}$, and~$\varphi_{\gamma}$ using~(\ref{eq.2.24}). If both of the roots~(\ref{eq.2.24}) are physical, one should perform in~(\ref{eq.B.9}) the summation over two values of~$E_{\ell}'$. In the limit when $m \ll E_{\ell}, E_{\ell}'$, the value of~$E_{\ell}'$ is unique and given by~(\ref{eq.2.26}), then the cross section~(\ref{eq.B.9}) can be written as
\begin{equation}
\frac{{\rm d} \sigma_{\text{brems}}}{{\rm d} E_{\gamma} \, {\rm d} \Omega_{\gamma} \, {\rm d} \Omega_{\ell}} = \frac{1}{(4\pi)^5} \, \frac{1}{M |{\boldsymbol \ell}|} \, \frac{E_{\gamma} \bigl[M \left(E_{\ell} - E_{\gamma}\right) - E_{\ell} E_{\gamma} \left(1 - \cos{\theta_{\gamma}}\right)\bigr]}{\bigl[M + E_{\ell} \left(1 - \cos{\theta_{\ell}}\right) - E_{\gamma} \left(1 - \cos{\psi}\right)\bigr]^2} \, |{\mathcal M}_{\text{brems}}|^2. \label{eq.B.10}
\end{equation}

\label{ref.1}
\pdfbookmark[1]{References}{ref.1}


\begin{thebibliography}{99}
\small
\bibitem{PRL.84.1398}
\emph{M.\,K.~Jones, K.\,A.~Aniol, F.\,T.~Baker, et~al.} $G_{E_p} / G_{M_p}$ ratio by polarization transfer in $\vec e p \rightarrow e \vec p$. Phys. Rev. Lett. \textbf{84} (2000) \href{http://dx.doi.org/10.1103/PhysRevLett.84.1398}{1398--1402}, arXiv:nucl-ex/\href{http://arxiv.org/abs/nucl-ex/9910005}{9910005}.

\bibitem{PRC.71.055202}
\emph{V.~Punjabi, C.\,F.~Perdrisat, K.\,A.~Aniol, et~al.} Proton elastic form factors ratios to $Q^2 = 3.5~\text{GeV}^2$ by polarization transfer. Phys. Rev. C \textbf{71} (2005) \href{http://dx.doi.org/10.1103/PhysRevC.71.055202}{055202}, arXiv:nucl-ex/\href{http://arxiv.org/abs/nucl-ex/0501018}{0501018}.

\bibitem{PRL.88.092301}
\emph{O.~Gayou, K.\,A.~Aniol, T.~Averett, et~al.} Measurement of $G_{E_p} / G_{M_p}$ in~$\vec e p \rightarrow e \vec p$ to $Q^2 = 5.6~\text{GeV}^2$. Phys. Rev. Lett. \textbf{88} (2002) \href{http://dx.doi.org/10.1103/PhysRevLett.88.092301}{092301}, arXiv:nucl-ex/\href{http://arxiv.org/abs/nucl-ex/0111010}{0111010}.

\bibitem{PRC.85.045203}
\emph{A.\,J.\,R.~Puckett, E.\,J.~Brash, O.~Gayou, et~al.} Final analysis of proton form factor ratio data at $Q^2 = 4.0$, $4.8$, and~$5.6~\text{GeV}^2$. Phys. Rev. C \textbf{85} (2012) \href{http://dx.doi.org/10.1103/PhysRevC.85.045203}{045203}, arXiv:\href{http://arxiv.org/abs/1102.5737}{1102.5737}.

\bibitem{PRL.104.242301}
\emph{A.\,J.\,R.~Puckett, E.\,J.~Brash, M.\,K.~Jones, et~al.} Recoil polarization measurements of the proton electromagnetic form factor ratio to~$Q^2 = 8.5~\text{GeV}^2$. Phys. Rev. Lett. \textbf{104} (2010) \href{http://dx.doi.org/10.1103/PhysRevLett.104.242301}{242301}, arXiv:\href{http://arxiv.org/abs/1005.3419}{1005.3419}.

\bibitem{PPNP.59.694}
\emph{C.\,F.~Perdrisat, V.~Punjabi, M.~Vanderhaeghen.} Nucleon electromagnetic form factors. Prog. Part. Nucl. Phys. \textbf{59} (2007) \href{http://dx.doi.org/10.1016/j.ppnp.2007.05.001}{694--764}, arXiv:hep-ph/\href{http://arxiv.org/abs/hep-ph/0612014}{0612014}.

\bibitem{Nature.466.213}
\emph{R.~Pohl, A.~Antognini, F.~Nez, et~al.} The size of the proton. Nature \textbf{466} (2010) \href{http://dx.doi.org/10.1038/nature09250}{213--216}.

\bibitem{Science.339.417}
\emph{A.~Antognini, F.~Nez, K.~Schuhmann, et~al.} Proton structure from the measurement of 2S-2P transition frequencies of muonic hydrogen. Science \textbf{339} (2013) \href{http://dx.doi.org/10.1126/science.1230016}{417--420}.

\bibitem{RMP.80.633}
\emph{P.\,J.~Mohr, B.\,N.~Taylor, D.\,B.~Newell.} CODATA recommended values of the fundamental physical constants: 2006. Rev. Mod. Phys. \textbf{80} (2008) \href{http://dx.doi.org/10.1103/RevModPhys.80.633}{633--730}, arXiv:\href{http://arxiv.org/abs/arXiv:0801.0028}{0801.0028}.

\bibitem{PRL.105.242001}
\emph{J.\,C.~Bernauer, P.~Achenbach, C.~Ayerbe~Gayoso, et~al.} High-precision determination of the electric and magnetic form factors of the proton. Phys. Rev. Lett. \textbf{105} (2010) \href{http://dx.doi.org/10.1103/PhysRevLett.105.242001}{242001}, arXiv:\href{http://arxiv.org/abs/1007.5076}{1007.5076}.

\bibitem{PLB.705.59}
\emph{X.~Zhan, K.~Allada, D.\,S.~Armstrong, et~al.} High-precision measurement of the proton elastic form factor ratio $\mu_p G_E / G_M$ at low~$Q^2$. Phys. Lett. B \textbf{705} (2011) \href{http://dx.doi.org/10.1016/j.physletb.2011.10.002}{59--64}, arXiv:\href{http://arxiv.org/abs/1102.0318}{1102.0318}.

\bibitem{ARNPS.63.175}
\emph{R.~Pohl, R.~Gilman, G.\,A.~Miller, K.~Pachucki.} Muonic hydrogen and the proton radius puzzle. Annu. Rev. Nucl. Part. Sci. \textbf{63} (2013) \href{http://dx.doi.org/10.1146/annurev-nucl-102212-170627}{175--204}, arXiv:\href{http://arxiv.org/abs/1301.0905}{1301.0905}.

\bibitem{ARNPS.57.171}
\emph{C.\,E.~Carlson, M.~Vanderhaeghen.} Two-photon physics in hadronic processes. Annu. Rev. Nucl. Part. Sci. \textbf{57} (2007) \href{http://dx.doi.org/10.1146/annurev.nucl.57.090506.123116}{171--204}, arXiv:hep-ph/\href{http://arxiv.org/abs/hep-ph/0701272}{0701272}.

\bibitem{PPNP.66.782}
\emph{J.~Arrington, P.\,G.~Blunden, W.~Melnitchouk.} Review of two-photon exchange in electron scattering. Prog. Part. Nucl. Phys. \textbf{66} (2011) \href{http://dx.doi.org/10.1016/j.ppnp.2011.07.003}{782--833}, arXiv:\href{http://arxiv.org/abs/1105.0951}{1105.0951}.

\bibitem{PRD.49.5671}
\emph{R.\,C.~Walker, B.\,W.~Filippone, J.~Jourdan, et~al.} Measurements of the proton elastic form factors for $1 \le Q^2 \le 3 \; (\text{GeV}/c)^2$ at SLAC. Phys. Rev. D \textbf{49} (1994) \href{http://dx.doi.org/10.1103/PhysRevD.49.5671}{5671--5689}.

\bibitem{PRL.94.142301}
\emph{I.\,A.~Qattan, J.~Arrington, R.\,E.~Segel, et~al.} Precision Rosenbluth measurement of the proton elastic form factors. Phys. Rev. Lett. \textbf{94} (2005) \href{http://dx.doi.org/10.1103/PhysRevLett.94.142301}{142301}, arXiv:nucl-ex/\href{http://arxiv.org/abs/nucl-ex/0410010}{0410010}.

\bibitem{PRC.70.068202}
\emph{J.\,J.~Kelly.} Simple parametrization of nucleon form factors. Phys. Rev. C \textbf{70} (2004) \href{http://dx.doi.org/10.1103/PhysRevC.70.068202}{068202}.

\bibitem{arXiv:1008.0855}
\emph{A.\,J.\,R.~Puckett.} Final results of the \mbox{GEp-III} experiment and the status of the proton form factors. arXiv:\href{http://arxiv.org/abs/1008.0855}{1008.0855}.

\bibitem{nucl-ex.0408020}
\emph{J.~Arrington, V.\,F.~Dmitriev, R.\,J.~Holt, et~al.} Two-photon exchange and elastic scattering of electrons/positrons on the proton. arXiv:nucl-ex/\href{http://arxiv.org/abs/nucl-ex/0408020}{0408020}.

\bibitem{NPBPS.225-227.216}
\emph{A.\,V.~Gramolin, J.~Arrington, L.\,M.~Barkov, et~al.} Measurement of the two-photon exchange contribution in elastic $ep$~scattering at \mbox{VEPP--3}. Nucl. Phys. B (Proc. Suppl.) \textbf{225--227} (2012) \href{http://dx.doi.org/10.1016/j.nuclphysbps.2012.02.045}{216--220}, arXiv:\href{http://arxiv.org/abs/1112.5369}{1112.5369}.

\bibitem{AIPConfProc.1160.19}
\emph{M.~Kohl.} The OLYMPUS experiment at DESY. AIP Conf. Proc. \textbf{1160} (2009) \href{http://dx.doi.org/10.1063/1.3232027}{19--23}.

\bibitem{NIMA.741.1}
\emph{R.~Milner, D.\,K.~Hasell, M.~Kohl, et~al.} The OLYMPUS experiment. Nucl. Instrum. Meth. A \textbf{741} (2014) \href{http://dx.doi.org/10.1016/j.nima.2013.12.035}{1--17}, arXiv:\href{http://arxiv.org/abs/1312.1730}{1312.1730}.

\bibitem{AIPConfProc.1160.24}
\emph{L.\,B.~Weinstein.} Electron- and positron--proton elastic scattering in CLAS. AIP Conf. Proc. \textbf{1160} (2009) \href{http://dx.doi.org/10.1063/1.3232028}{24--28}.

\bibitem{PRC.88.025210}
\emph{M.~Moteabbed, M.~Niroula, B.\,A.~Raue, et~al.} Demonstration of a novel technique to measure two-photon exchange effects in elastic $e^{\pm} p$~scattering. Phys. Rev. C \textbf{88} (2013) \href{http://dx.doi.org/10.1103/PhysRevC.88.025210}{025210}, arXiv:\href{http://arxiv.org/abs/1306.2286}{1306.2286}.

\bibitem{PRad}
\emph{A.~Gasparian, R.~Pedroni, Z.~Ahmed, et~al.} High precision measurement of the proton charge radius. Jefferson Lab proposal C12-11-106 (\url{http://www.jlab.org/exp_prog/proposals/12/C12-11-106.pdf}).

\bibitem{EPJWebConf.73.07006}
\emph{A.~Gasparian.} The PRad experiment and the proton radius puzzle. EPJ Web of Conferences \textbf{73} (2014) \href{http://dx.doi.org/10.1051/epjconf/20147307006}{07006}.

\bibitem{arXiv:1303.2160}
\emph{R.~Gilman, E.\,J.~Downie, G.~Ron, et~al.} Studying the proton ``radius'' puzzle with $\mu p$~elastic scattering. arXiv:\href{http://arxiv.org/abs/1303.2160}{1303.2160}.

\bibitem{arXiv:1207.5089}
\emph{B.~Wojtsekhowski, D.~Nikolenko, I.~Rachek.} Searching for a new force at \mbox{VEPP--3}. arXiv:\href{http://arxiv.org/abs/1207.5089}{1207.5089}.

\bibitem{PRD.86.115012}
\emph{Y.~Kahn, J.~Thaler.} Searching for an invisible $A'$ vector boson with DarkLight. Phys. Rev. D \textbf{86} (2012) \href{http://dx.doi.org/
10.1103/PhysRevD.86.115012}{115012}, arXiv:\href{http://arxiv.org/abs/arXiv:1209.0777}{1209.0777}.

\bibitem{arXiv:1307.4432}
\emph{J.~Balewski, J.~Bernauer, W.~Bertozzi, et~al.} DarkLight: a search for dark forces at the Jefferson Laboratory Free-Electron Laser facility. arXiv:\href{http://arxiv.org/abs/1307.4432}{1307.4432}.

\bibitem{PR.122.1898}
\emph{Y.-S. Tsai.} Radiative corrections to electron--proton scattering. Phys. Rev. \textbf{122} (1961) \href{http://dx.doi.org/10.1103/PhysRev.122.1898}{1898--1907}.

\bibitem{RMP.41.205}
\emph{L.\,W.~Mo, Y.\,S.~Tsai.} Radiative corrections to elastic and inelastic $ep$ and $\mu p$ scattering. Rev. Mod. Phys. \textbf{41} (1969) \href{http://dx.doi.org/10.1103/RevModPhys.41.205}{205--235}.

\bibitem{PRC.62.054320}
\emph{L.\,C.~Maximon, J.\,A.~Tjon.} Radiative corrections to electron--proton scattering. Phys. Rev. C \textbf{62} (2000) \href{http://dx.doi.org/10.1103/PhysRevC.62.054320}{054320}, arXiv:nucl-th/\href{http://arxiv.org/abs/nucl-th/0002058}{0002058}.

\bibitem{PRC.64.054610}
\emph{R.~Ent, B.\,W.~Filippone, N.\,C.\,R.~Makins, et~al.} Radiative corrections for $(e, e'p)$ reactions at GeV energies. Phys. Rev. C \textbf{64} (2001) \href{http://dx.doi.org/10.1103/PhysRevC.64.054610}{054610}.

\bibitem{PR.130.1210}
\emph{N.~Meister, D.\,R.~Yennie.} Radiative corrections to high-energy scattering processes. Phys. Rev. \textbf{130} (1963) \href{http://dx.doi.org/10.1103/PhysRev.130.1210}{1210--1229}.

\bibitem{PRC.62.025501}
\emph{M.~Vanderhaeghen, J.\,M.~Friedrich, D.~Lhuillier, et~al.} QED radiative corrections to virtual Compton scattering. Phys. Rev. C \textbf{62} (2000) \href{http://dx.doi.org/10.1103/PhysRevC.62.025501}{025501}, arXiv:hep-ph/\href{http://arxiv.org/abs/hep-ph/0001100}{0001100}.

\bibitem{PRD.64.113009}
\emph{A.~Afanasev, I.~Akushevich, N.~Merenkov.} Model independent radiative corrections in processes of polarized electron-nucleon elastic scattering. Phys. Rev. D \textbf{64} (2001) \href{http://dx.doi.org/10.1103/PhysRevD.64.113009}{113009}, arXiv:hep-ph/\href{http://arxiv.org/abs/hep-ph/0102086}{0102086}.

\bibitem{PRC.75.015207}
\emph{Yu.\,M.~Bystritskiy, E.\,A.~Kuraev, E.~Tomasi-Gustafsson.} Structure function method applied to polarized and unpolarized electron--proton scattering: a solution of the $G_E(p) / G_M(p)$ discrepancy. Phys. Rev. C \textbf{75} (2007) \href{http://dx.doi.org/10.1103/PhysRevC.75.015207}{015207}, arXiv:hep-ph/\href{http://arxiv.org/abs/hep-ph/0603132}{0603132}.

\bibitem{arXiv:1311.0370}
\emph{E.\,A.~Kuraev, A.\,I.~Ahmadov, Yu.\,M.~Bystritskiy, E.\,Tomasi-Gustafsson.} Radiative corrections for electron proton elastic scattering taking into account high orders and hard photon emission. arXiv:\href{http://arxiv.org/abs/1311.0370}{1311.0370}.

\bibitem{AnnPhys.13.379}
\emph{D.\,R.~Yennie, S.\,C.~Frautschi, H.~Suura.} The infrared divergence phenomena and high-energy processes. Annals of Physics \textbf{13} (1961) \href{http://dx.doi.org/10.1016/0003-4916(61)90151-8}{379--452}.

\bibitem{PR.79.615}
\emph{M.\,N.~Rosenbluth.} High energy elastic scattering of electrons on protons. Phys. Rev. \textbf{79} (1950) \href{http://dx.doi.org/10.1103/PhysRev.79.615}{615--619}.

\bibitem{PRC.36.2466}
\emph{B.\,M.~Preedom, R.~Tegen.} Nucleon electromagnetic form factors from scattering of polarized muons or electrons. Phys. Rev. C \textbf{36} (1987) \href{http://dx.doi.org/10.1103/PhysRevC.36.2466}{2466--2472}.

\bibitem{PRC.42.599}
\emph{P.\,C.~Tiemeijer, J.\,A.~Tjon.} Electromagnetic form factors for an off-shell nucleon in a vector meson dominance model. Phys. Rev. C \textbf{42} (1990) \href{http://dx.doi.org/10.1103/PhysRevC.42.599}{599--609}.

\bibitem{FeynCalc}
\url{http://www.feyncalc.org}

\bibitem{ESEPP}
\url{https://github.com/gramolin/esepp/}

\bibitem{Berestetskii}
\emph{V.\,B.~Berestetskii, E.\,M.~Lifshitz, L.\,P.~Pitaevskii.} Quantum Electrodynamics (Course of Theoretical Physics, Vol.~4). Pergamon, Oxford, 1982.

\bibitem{vpl}
\url{http://cmd.inp.nsk.su/~ignatov/vpl/}

\bibitem{EPJC.66.585}
\emph{S.~Actis, A.~Arbuzov, G.~Balossini, et~al.} Quest for precision in hadronic cross sections at low energy: Monte Carlo tools vs. experimental data. Eur. Phys. J. C \textbf{66} (2010) \href{http://dx.doi.org/10.1140/epjc/s10052-010-1251-4}{585--686}, arXiv:\href{http://arxiv.org/abs/0912.0749}{0912.0749}.

\bibitem{arXiv:1207.6651}
\emph{E.~Borie.} Muon-proton scattering. arXiv:\href{http://arxiv.org/abs/1207.6651}{1207.6651}.

\bibitem{ROOT}
\url{http://root.cern.ch}

\bibitem{CPC.152.55}
\emph{S.~Jadach.} Foam: a general-purpose cellular Monte Carlo event generator. Comput. Phys. Commun. \textbf{152} (2003) \href{http://dx.doi.org/10.1016/S0010-4655(02)00755-5}{55--100}, arXiv:physics/\href{http://arxiv.org/abs/physics/0203033}{0203033}.

\bibitem{CPC.177.441}
\emph{S.~Jadach, P.~Sawicki.} \mbox{mFOAM-1.02}: a compact version of the cellular event generator FOAM. Comput. Phys. Commun. \textbf{177} (2007) \href{http://dx.doi.org/10.1016/j.cpc.2007.02.112}{441--458}, arXiv:physics/\href{http://arxiv.org/abs/physics/0506084}{0506084}.

\bibitem{Itzykson&Zuber}
\emph{C.~Itzykson, J.-B.~Zuber.} Quantum Field Theory. McGraw-Hill, New York, 1980.
\end{thebibliography}
\end{document}